\documentclass[aip,preprint,amsmath,amssymb,floatfix,nofootinbib]{revtex4-1}

\usepackage{graphicx}
\usepackage{amsmath}
\usepackage{hyperref}

\begin{document}

\title{Single beam collective effects in FCC-ee due to beam coupling impedance}

%\vspace{-.3truecm}
\author{E. Belli}
\affiliation{University of Rome 'La Sapienza', INFN Sez. Roma1 - Roma - Italy, and CERN, Geneva, Switzerland}
\author{M. Migliorati}
\email{mauro.migliorati@uniroma1.it}
\affiliation{University of Rome 'La Sapienza' and INFN Sez. Roma1 - Roma - Italy}
\author{S. Persichelli}
\affiliation{LBNL - Berkeley - CA - USA}
\author{M. Zobov}
\affiliation{INFN-LNF - Frascati - Roma - Italy}

\date{\today}

\begin{abstract}
The Future Circular Collider study, hosted by CERN to design post-LHC particle accelerator options in a worldwide context, is focused on proton-proton high-energy and electron-positron high-luminosity frontier machines. This new accelerator complex represents a great challenge under several aspects, which involve R\&D on beam dynamics and new technologies. One very critical point in this context is represented by collective effects, generated by the interaction of the beam with self-induced electromagnetic fields, called wake fields, which could produce beam instabilities, thus reducing the machines performance and limiting the maximum stored current. It is therefore very important to be able to predict these effects and to study in detail potential solutions to counteract them. In this paper the resistive wall and some other important geometrical sources of impedance for the FCC electron-positron accelerator are identified and evaluated, and their impact on the beam dynamics, which in some cases could lead to unwanted instabilities, is discussed.
\end{abstract}

\pacs{29.20.db, 29.27.Bd, 41.75.Ht }% insert suggested PACS numbers in braces
% PACS, the Physics and Astronomy Classification Scheme.
\keywords{Beam Dynamics, collective effects, wakefields, instabilities.}%Use showkeys class option if keyword
                              %display desired

\maketitle %\maketitle must follow title, authors, abstract and \pacs

\section{Introduction}
The Large Hadron Collider (LHC) at CERN, is nowadays at the start of a new program which is expected to run for about other 20 years. It is called High Luminosity LHC~\cite{HL-LHC} (HL-LHC) and it is CERN's number-one priority which aims to increase the number of collisions accumulated in the experiments by a factor of ten from 2024 onwards.

While HL-LHC project is already well defined for the next two decades, CERN has started an exploratory study for a future long-term project based on a new generation of circular colliders with a circumference of about 100 km. The Future Circular Collider (FCC) study~\cite{FCC}  has been undertaken to design a high energy proton-proton machine (FCC-hh), capable of reaching unprecedented energies in the region of 100 TeV, and a high-luminosity e+e- collider (FCC-ee), serving as $Z$, $W$, Higgs and top factory, with luminosities ranging from about $10^{34}$ to $10^{36}$ cm$^{-2}$s$^{-1}$ per collision point as a potential intermediate step towards the realization of the hadron facility. The design of the lepton collider complex will be based on the same infrastructure as the hadron collider.

At high beam intensity, necessary to reach the high luminosity foreseen for FCC-ee, the electromagnetic fields, self-generated by the beam interacting with its immediate surroundings and known as wake fields~\cite{palumbo}, act back on the beam, perturbing the external guiding fields and the beam dynamics. Under unfavorable conditions, the perturbation on the beam further enhances the wake fields; the beam-surroundings interaction then can lead to a reduction of the machine performances and, in some cases, also to instabilities.

The theory of collective beam instabilities induced by the wake fields is a broad subject and it has been assessed over many years by the work of several authors, such as F.~Sacherer~\cite{sacherer}, A.~W.~Chao~\cite{chao}, J.~L.~Laclare~\cite{laclare}, B.~Zotter~\cite{zotter}, C.~Pellegrini~\cite{pellegrini}, M.~Sands~\cite{sands} and others~\cite{yokoya1}.

To simplify the  study of collective effects, in general it is convenient to distinguish between short range wake fields, which influence the single bunch beam dynamics, and long range wake fields, where high quality factor resonant modes excited by a train of bunches can last for many turns exciting, under some conditions, coupled bunch instabilities. In both cases the bunch motion is considered as a sum of coherent oscillation modes perturbed by these wake fields. 

In this paper we will focus on the FCC-ee collective effects induced by wake fields. In particular we will first evaluate the wake fields induced by the finite resistivity of the beam vacuum chamber (resistive wall) in section~\ref{s:sec1}. Due to the 100 km of length of the beam pipe, the resistive wall is going to play a fundamental role among the sources of wake fields for this accelerator, and the choice of the pipe geometry (circular or elliptical), material, and dimensions is particularly important.  In section~\ref{s:sec2} we study the collective effects induced by the resistive wall for both the short range (\ref{s:sub1}) and long range wake fields (\ref{s:sub2}), and for both longitudinal and transverse planes. For some instabilities we will resort to the linear theory, while for other cases and for more accurate predictions, we need to use simulation codes.

Section~\ref{s:sec3} is then dedicated to other important sources of wake fields, such as the RF system and tapers, and to the discussions on the choice of some devices, as the synchrotron radiation absorbers, in order to reduce their impact on the total wake field. Finally, concluding remarks and outlook will end the paper.

For reference we report in Table~\ref{tab1} the list of beam parameters we have used for evaluating the effects of wake fields on the beam dynamics.

\begin{table}[!ht]
\caption{Parameter list used to evaluate the beam dynamics effects of wake fields.}
\begin{center}
\begin{tabular}{|l|c|c|c|c|}
\hline
Circumference (km)&  100 & 100 &  100 & 100 \\
\hline
Beam energy (GeV)&  45.6 & 80 &  120 & 175 \\
\hline
Beam current (mA)&  1450 & 152 &  30 & 6.6 \\
\hline
RF frequency (MHz)&  400 & 400 &  400 & 400 \\
\hline
RF voltage (GV)&  0.2 & 0.8 &  3 & 10 \\
\hline
Mom.~compaction (10$^{-5}$) &  0.7 & 0.7 &  0.7 & 0.7 \\
\hline
Bunch length (mm)* &  1.6 & 2.0 &  2.0 & 2.1 \\
\hline
Energy spread (10$^{-3}$)* &  0.37 & 0.65 &  1.0 & 1.4 \\
\hline
Synchrotron tune &  0.025 & 0.037 &  0.056 & 0.075 \\
\hline
Bunches/beam &  90300 & 5260 &  780 & 81 \\
\hline
Bunch population (10$^{11}$)&  0.33 & 0.6 &  0.8 & 1.7 \\
\hline
Betatron tune &  350 & 350 &  350 & 350 \\
\hline
\end{tabular}
\\ * without beamstrahlung (no collision, worst case)
\end{center}
\label{tab1}
\end{table}%

\section{Resistive wall wake fields and impedances}
\label{s:sec1}

The electromagnetic interaction of the beam with the surrounding vacuum chamber, due to its finite resistivity, produces unavoidable wake fields, which, for FCC-ee, result to be of particular importance. If we consider a beam pipe with circular cross section and a single material of infinite thickness, the longitudinal monopolar ($m=0$) coupling impedance is given by~\cite{bane}
\begin{equation}
\label{rw1}
\frac{Z_{||} \left(\omega\right)}{C}=\frac{Z_0 c}{\pi} \frac{1}{\left[1+i\textrm{sgn}\left(\omega\right)\right] 2bc
\sqrt{\frac{\sigma_c Z_0 c}{2 \left| \omega \right|}}-i b^2 \omega}
\end{equation}
and the transverse dipolar ($m=1$) one by
\begin{equation}
\label{rw2}
\frac{Z_{\perp} \left(\omega\right)}{C}=\frac{Z_0 c^2}{\pi} \frac{2}{\left[\textrm{sgn}\left(\omega\right)+i\right] b^3c
\sqrt{2\sigma_c Z_0 c\left| \omega \right|}-i b^4 \omega^2}
\end{equation}
where $C$ is the machine circumference, $Z_0$ the vacuum impedance, $c$ the speed of light, $b$ the pipe radius, and $\sigma_c$ the material conductivity. The above expressions are valid in a frequency range defined by
\begin{equation}
\frac{\chi c} {b} \ll \omega \ll \frac{c \chi^{-1/3}}{b}
\end{equation}
with $\chi=1/\left(Z_0 \sigma_c c b \right)$. The corresponding wake functions are given by~\cite{bane1}
\begin{equation}
\label{eq:long_wake}
\frac{w_{||}\left( z \right)}{C} =\frac{4Z_0 c}{\pi b^2} \left[ \frac{e^{-z/s_0}}{3} 
\cos \left( \frac{\sqrt{3}z}{s_0} \right) 
-\frac{\sqrt{2}}{\pi} \int_{0}^{\infty} dx \frac{x^2 e^{-z x^2/s_0}}{x^6+8}
\right]
\end{equation}
and
\begin{equation}
w_{\perp}\left( z \right)=\frac{2} {b^2} \frac{dw_{||}\left( z \right)}{dz}
\end{equation}
with $z>0$ and $s_0=\left[2b^2/\left(Z_0 \sigma_c \right) \right]^{1/3}$.

By considering a beam pipe of 35 mm of radius made by copper (conductivity of about 5.9e7 S/m) or aluminium (conductivity of about 3.8e7 S/m), eqs.~(\ref{rw1}) and (\ref{rw2}) are valid in a very large range of frequency. 
In addition, it is important to observe that the last term in the denominator of eqs.~(\ref{rw1}) and (\ref{rw2}) is negligible up to high frequencies, giving then the possibility to easily evaluate the scale of the impedance with the pipe radius. Indeed the longitudinal impedance is inversely proportional to the beam pipe radius, and the transverse one to the inverse of the third power of $b$. This scaling can be used to find a compromise for the pipe geometry. By reducing the radius it is possible to reduce the power required for the magnets, but this would increase in particular the coupling impedance and then reduce transverse instability thresholds.

In the following we will use, as a value for the pipe radius, 35 mm, which, as it will be shown in the next section, gives instability thresholds which are quite safe with respect to the nominal beam parameters\footnote{Except for the transverse coupled bunch instability for which, as it will be discussed in section~\ref{s:sub2}, an active feedback system has to be foreseen.}, and represents a good compromise also for the quadrupole magnets' power. Indeed, by using the impedance scaling with radius it is easy to infer scenarios at different pipe radii.

The above equations represent the impedance per unit of length. Due to the fact that the length of the accelerator in this case is of unprecedented value, also the total resistive wall coupling impedance is very high. However it is important to note that the effects of the impedance on the beam dynamics do not scale directly with the machine length. As we will see in the next section, other parameters peculiar of this machine, as the energy and the tunes, are important. In addition we also observe that the above impedances depend on the frequency where they are evaluated. If, for example, we consider the transverse impedance at the revolution frequency, we see that it approximately scales as $C\sqrt{C}$, and not simply with $C$. Some discussions about the scales will be done when dealing with the different instabilities in the next section, when we will see that the mere contribution of the resistive wall to the total wake fields gives important effects in both the longitudinal and transverse planes.

Concerning the choice of the material, we see that, since both the impedances are proportional to the square root of the resistivity, the difference between copper and aluminium gives a factor of about 1.26, which is not considered a critical value. 

\begin{figure}[!ht]
\center
\includegraphics[width=115mm]{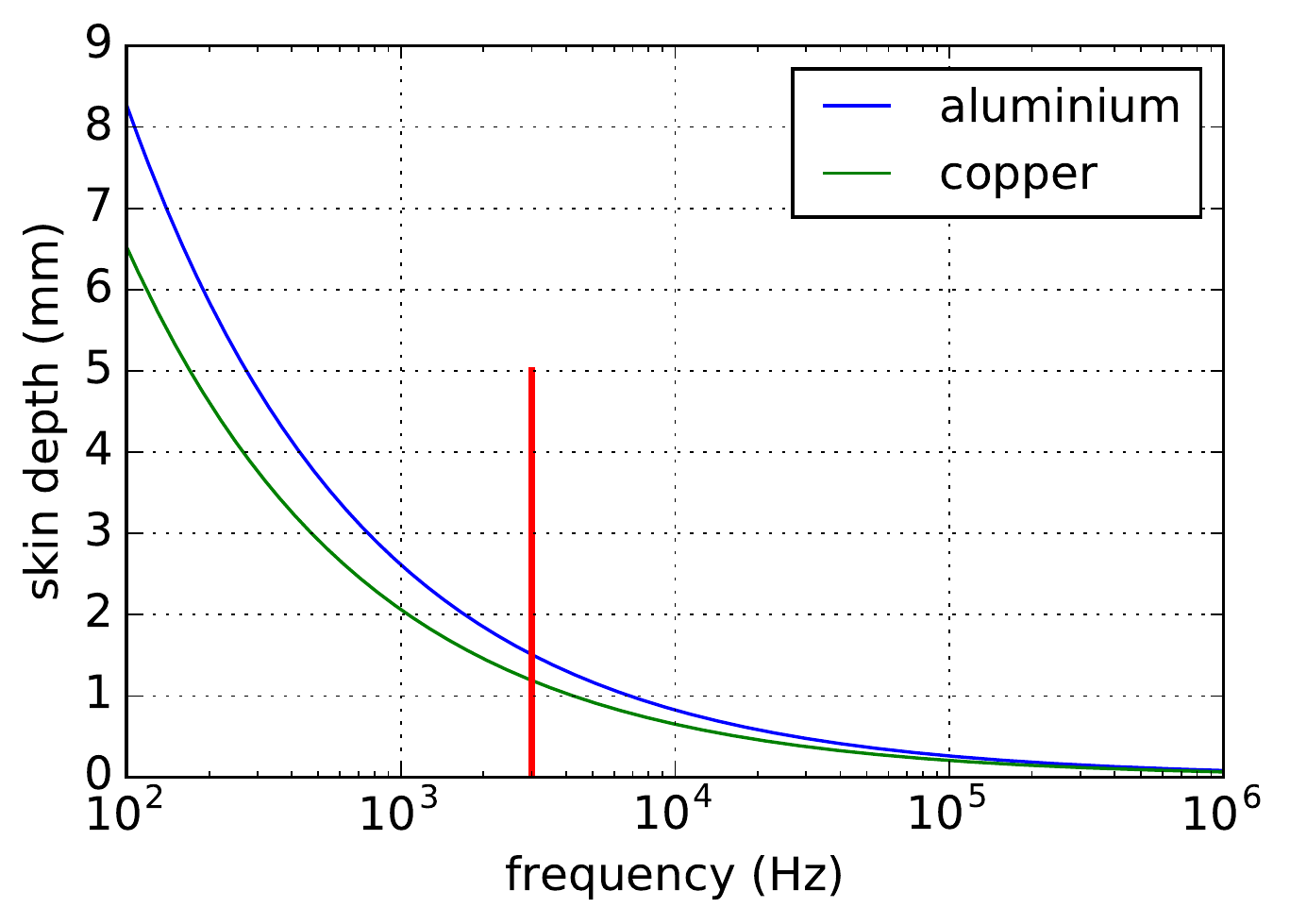}
  \caption{Skin depth as a function of frequency for aluminium (blue) and copper (green). The red line represents the revolution frequency.}
\label{skindepth}
\end{figure}

The above equations, however, are valid only for a single thick layer, for which the skin depth is much smaller than the wall thickness. In Fig.~\ref{skindepth} we show the skin depth for copper and aluminium as a function of frequency. We can see from the plot that at the revolution frequency of about 3 kHz (red line in the figure), the skin depths are about 1.2 and 1.5 mm. For lower frequencies the skin depth is higher. At the moment of writing this paper, the first layer of the vacuum chamber is aluminium with a thickness of 4 mm~\cite{kersevan1}, that means that eqs.~(\ref{rw1}) and (\ref{rw2}) can be used up to very low frequencies, and only at about 100 Hz the difference between one and multiple layers starts to give some differences.

\begin{figure}[!ht]
\center
\includegraphics[width=115mm]{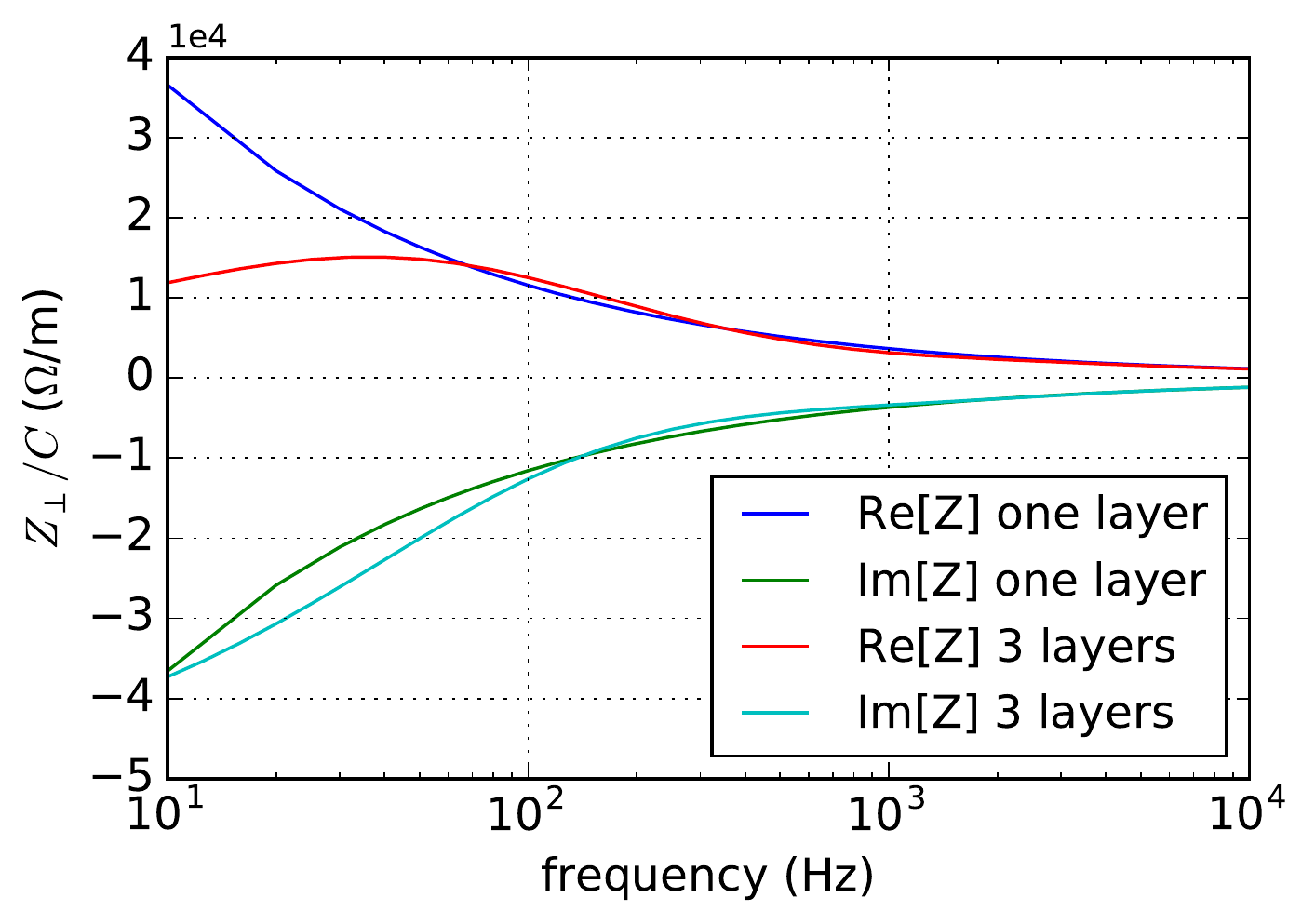}
  \caption{Transverse impedance as a function of frequency for single layer (real part in blue and imaginary part in green) and three layers (real part in red and imaginary part in cyan).}
\label{multilayer}
\end{figure}

In Fig.~\ref{multilayer} we show the transverse impedance $Z_{\perp} \left( \omega \right)/C$ as a function of frequency for a circular pipe with radius 35 mm, aluminium, in case of single layer, as given by eq.~(\ref{rw2}), and three layers, with a first layer of aluminium of 4 mm, then 6 mm of dielectric and finally iron with resistivity of $10^{-7}$ $\Omega$m. In the three layers case, the impedance has been evaluated with the code ImpedanceWake2D~\cite{mounet}. As shown in the figure the difference between the two impedances starts to show up at very low frequency. As a conclusion we can say that all the considerations derived from eqs.~(\ref{rw1}) and (\ref{rw2}) are essentially valid also for the multilayer case.

\begin{figure}[!ht]
\center
\includegraphics[width=115mm]{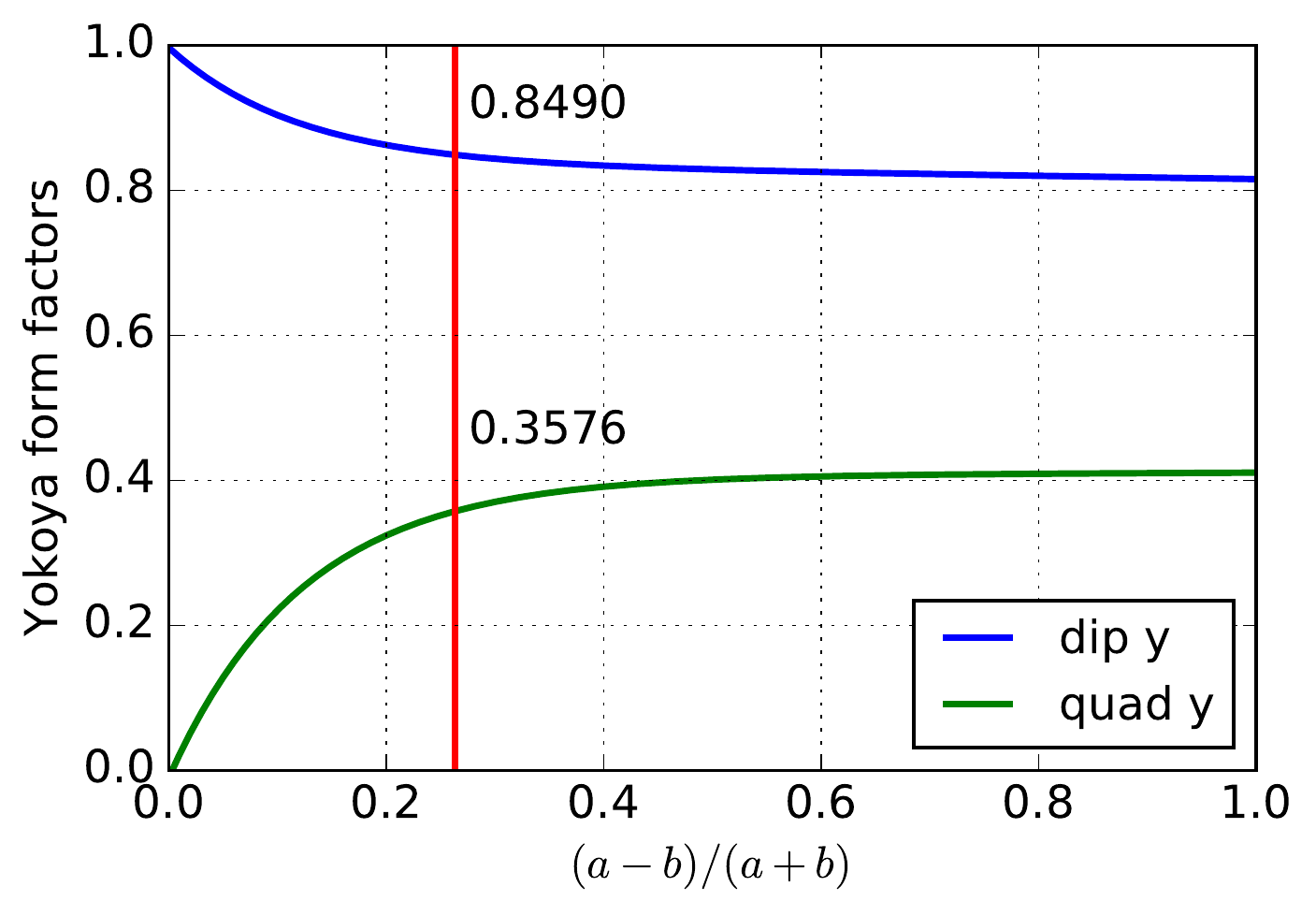}
  \caption{Yokoya form factors. $a$ and $b$ are the major and minor semi-axes respectively.}
\label{yokoya}
\end{figure}

For what concerns the pipe geometry, the options are between circular and elliptic cross section. An elliptic chamber with semi-axes 60x35 mm was initially considered for allowing to insert synchrotron radiation absorbers on one side of the vacuum chamber along the major semi-axis. However, even if the elliptic shape gives a reduced impedance in the horizontal plane, it produces an additional quadrupolar transverse impedance which has to be taken into account. In this case the Yokoya form factors~\cite{yokoya2}, shown in Fig.~\ref{yokoya}, give a reduction of the transverse impedance of about 15\%, but an additional quadrupolar impedance which is about 36\% the dipolar one.

\begin{figure}[!ht]
\center
\includegraphics[width=160mm]{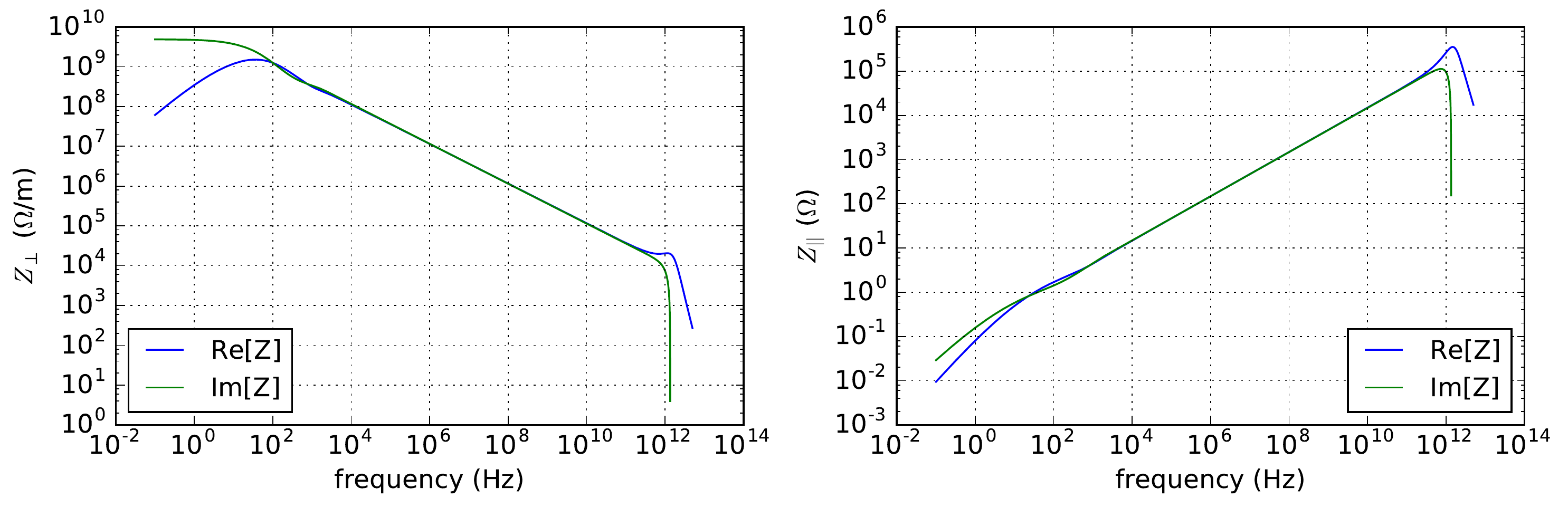}
  \caption{Real and imaginary part of transverse (left) and longitudinal (right) impedance of resistive wall as a function of frequency.}
\label{RW}
\end{figure}

Fig.~\ref{RW} shows the total transverse and longitudinal resistive wall impedance as a function of frequency for a circular beam pipe of 35 mm of radius in the three layer case. This impedance is used in the following section for evaluating the resistive wall effect on beam dynamics.

\section{Effects of resistive wall on beam dynamics}
\label{s:sec2}

\subsection{Single bunch effects}
\label{s:sub1}

\begin{figure}[!ht]
\center
\includegraphics[width=80mm]{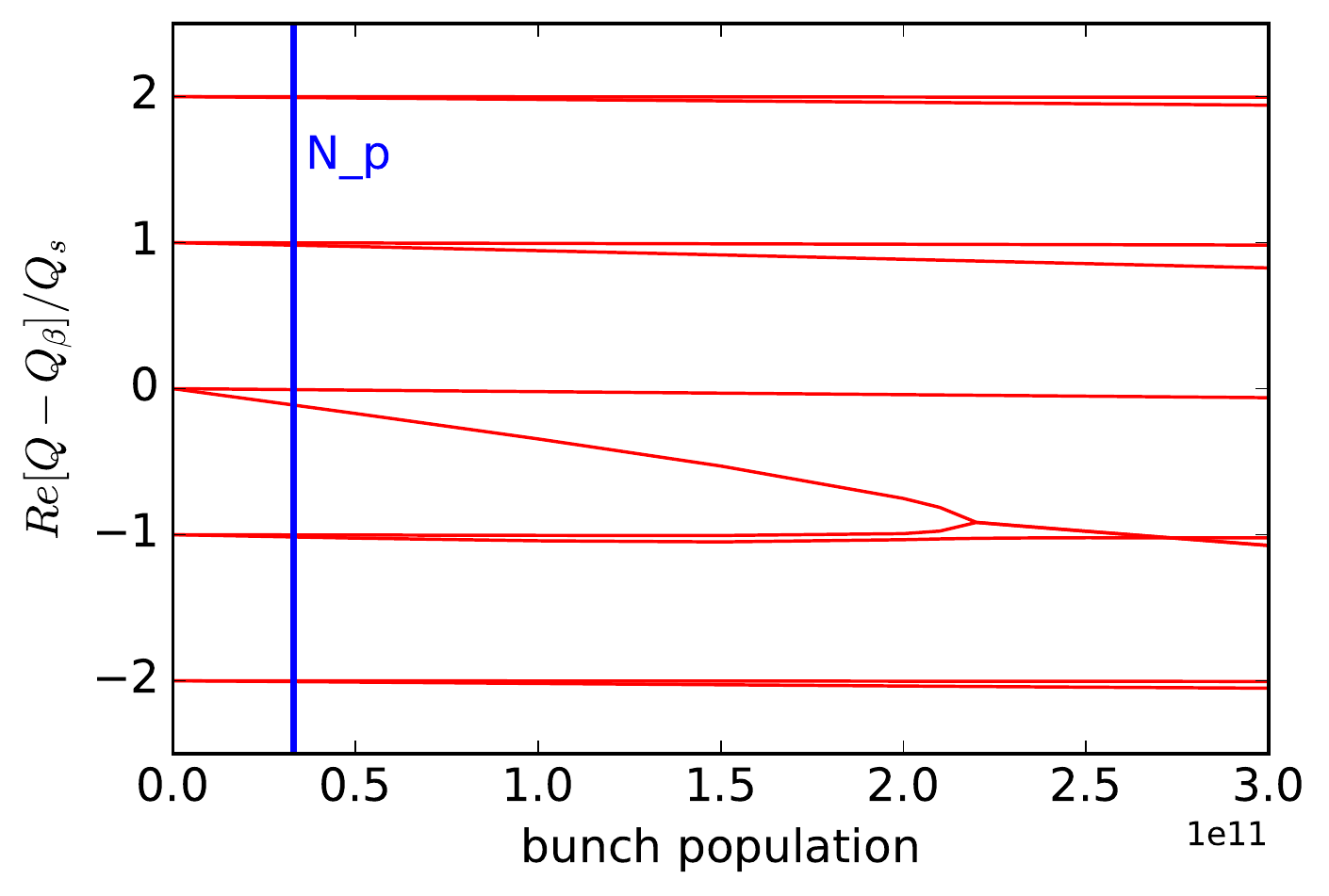}
\includegraphics[width=80mm]{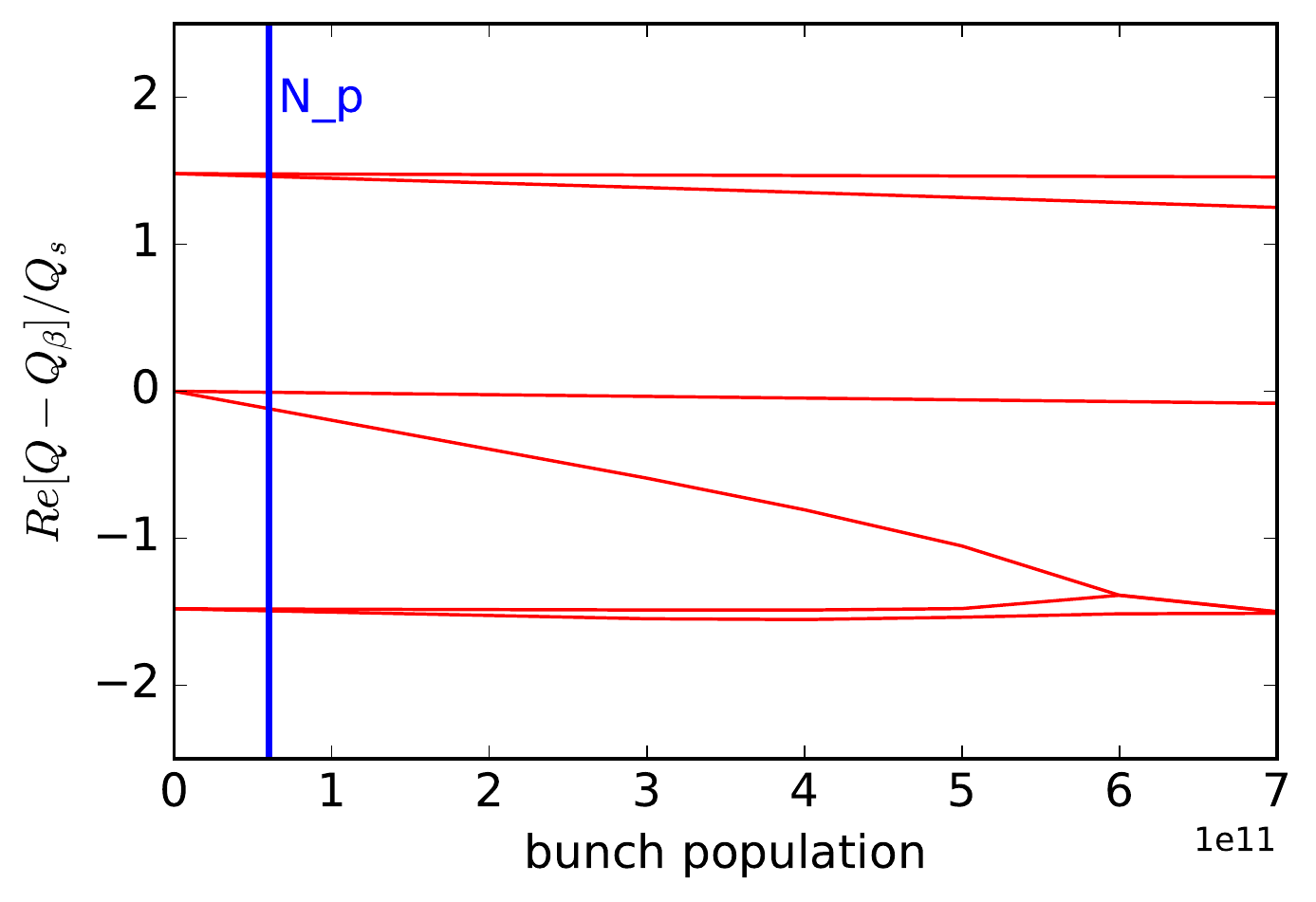}
   \caption{Real part of the frequency of the first coherent oscillation modes as a function of bunch population for the 45.6 GeV case (left) and the 80 GeV case (right).}
\label{TMCI}
\end{figure}

One important effect of the resistive wall on the single bunch dynamics is related to the transverse mode coupling instability, or strong head tail instability~\cite{chao}. In general the transverse motion of the bunch, derived from the Vlasov equation, can be decomposed as a sum of coherent modes of oscillation, called eigenmodes, the coherent frequencies of which depend on the current intensity and on the coupling impedance. If the bunch distribution function can be expressed as a sum of orthogonal polynomials, it is possible to evaluate amplitude and frequencies of the eigenmodes, that could couple together giving rise to an instability. The frequencies of the coherent modes are here calculated with DELPHI~\cite{mounet} code, which considers Laguerre polynomials. In Fig.~\ref{TMCI} we show the real part of the frequency (tune shift) of the first two radial coherent oscillation modes, with the azimuthal number going from -2 to 2, as a function of the bunch population for 45.6 GeV and 80 GeV. As expected, the worst scenario is at the lowest energy, where we find an instability threshold that is a factor of about 6 higher than the nominal bunch population. The higher energy cases, not shown here, give even higher thresholds. In this situation we can see that, if other contributions to the transverse impedance do not exceed the resistive wall, we have a good margin of safety for this kind of instability. However, a more detailed study of transverse mode coupling instability with a more detailed transverse impedance is necessary.

For what concerns the longitudinal beam dynamics, one main problem caused by the resistive wall is related to the longitudinal potential well distortion and the evaluation of the microwave instability threshold. The microwave instability does not produce a bunch loss, but the consequent longitudinal emittance increase and possible bunch internal oscillations that cannot be counteracted by a feedback system, make the microwave instability an effect that has to be studied with care. In addition to that, there are no reliable analytical expressions that can be used to easily evaluate the instability threshold. For these reasons we have performed a series of simulations by using a tracking code, which we refer here as SBSC~\cite{music}, initially developed to study the longitudinal beam dynamics in the electron storage ring DA$\Phi$NE~\cite{sbunch1}, and successively developed and adapted to other machines~\cite{sbunch2}.

\begin{figure}[!ht]
\center
\includegraphics[width=120mm]{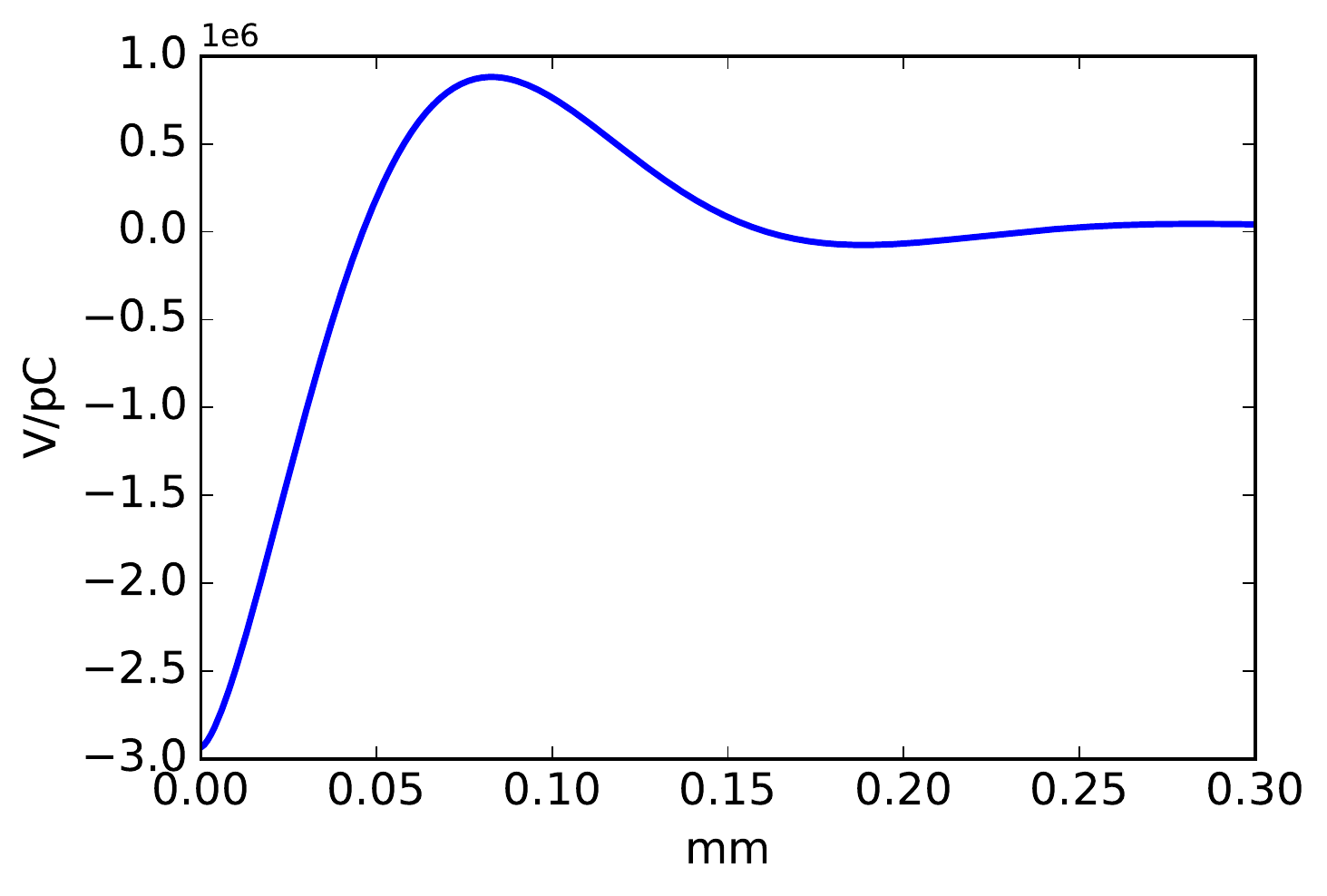}
   \caption{Resistive wall longitudinal wake function used for simulations.}
\label{long_wakefield}
\end{figure}

As other single bunch tracking codes taking into account the wake fields effects, in order to reduce the computing time, the code uses macro-particles, divides the bunch into slices, or bins, and evaluates the wake potential, that is the convolution integral of the wake function times the longitudinal distribution function, at the centre of each slice to then interpolate the wake for all the positions of the macro-particles. In Fig.~\ref{long_wakefield} we have represented the wake function given by eq.~(\ref{eq:long_wake}). As we can see, the wake varies considerably in a very short distance. This causes problems to simulations with slices. If we consider, for example, an rms bunch length $\sigma_z$ of about 3-4 mm, and use $\pm 5 \sigma_z$ for the beam dynamics, in order to have about 20 slices in 50 $\mu$m, we need to have several thousands of slices, making the simulation very time consuming.

\begin{figure}[!ht]
\center
\includegraphics[width=120mm]{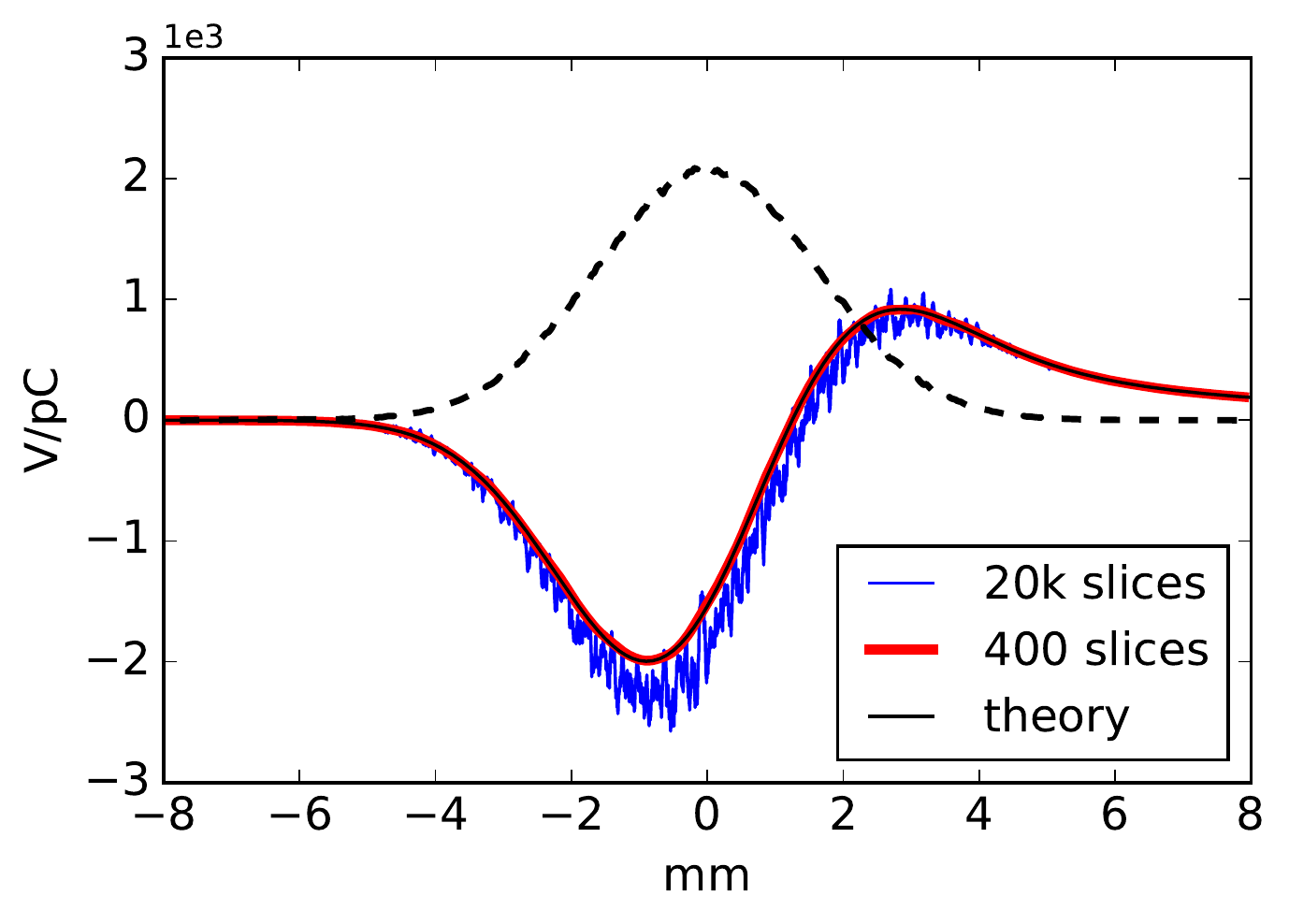}
   \caption{Comparison of the analytical longitudinal wake potential of a 1.6 mm Gaussian bunch (black line) with those given by the simulation code, one using $2\times 10^4$ slices and the wake function as Green function (blue line), and the second one with 400 slices and the wake potential of a 0.15 mm Gaussian bunch as Green function (red line).}
\label{long_wakepot_bins}
\end{figure}

In Fig.~\ref{long_wakepot_bins} in blue we show the wake potential of a 1.6 mm Gaussian bunch obtained with the tracking code using $2\times 10^4$ slices compared with the analytical calculation (black line) given by the equation~\cite{piwinski}
\begin{equation}
\label{eq:rw}
W_{||}(z)=\int_{-\infty}^{\infty}\lambda (z') w_{||}(z-z')dz'=\frac{cC}{8\sqrt{2}\pi b \sigma_z^{3/2}} \sqrt{\frac{Z_0}{\sigma_c}}F(z/\sigma_c)
\end{equation}
with
\begin{equation}
\label{bessel}
F(x)=|x|^{3/2} e^{-x^2/4}\left(I_{1/4}-I_{-3/4}\pm I_{-1/4} \mp I_{3/4} \right)
\end{equation}
where $I_n$ is the modified Bessel function, the upper signs in eq.~\ref{bessel} are for positive $z$, and $\lambda(z)$ is the longitudinal distribution function.

The oscillations of the wake from slice to slice given by the code are due to the fact that, in order to have a good statistics, in addition to the excessive number of slices, we should also use a very high number of macro-particles.

It is possible to overcome this problem by using not the wake function but the wake potential of a very short bunch as Green function for the code. The same figure, with the red line, shows the wake potential given by the same simulation code using only 400 slices and, as Green function, the wake potential of a 0.15 mm Gaussian bunch. As we see, the wake is much smoother due to the fact that the same number of macro-particles are now distributed over 400 slices instead on $2\times 10^4$, and the agreement with eq.~(\ref{eq:rw}) is excellent.

In order to test the code, and to evaluate the effect of the resistive wall on the longitudinal beam dynamics, we have first solved the Ha\"\i ssinski integral equation~\cite{haissinski}, which is able to predict the bunch length and the distortion from a Gaussian distribution for intensities below the microwave instability threshold. The equation can be written as
\begin{equation}
\lambda(z)=\lambda_0 \exp{ \left[ \frac{1} {E_0 \eta \sigma_{\varepsilon 0}^2} \Psi(z) \right]}
\end{equation}
with $\lambda_0$ a normalization constant, $E_0$ the collider energy, $\eta$ the slippage factor, $\sigma_{\varepsilon 0}$ the natural RMS energy spread, and
\begin{equation}
\Psi(z)=\frac{1}{C} \int_{0}^{z} \left[ e V_{RF}(z') - U_0\right] dz'-\frac{e^2 N_p} {C} \int_{0}^z dz' \int_{-\infty}^{z'} \lambda(z'') w_{||}(z'-z'') dz'' \ .
\end{equation}
where $V_{RF}$ represents the total RF voltage, $U_0$ the energy lost per turn due to the synchrotron radiation, and $N_p$ the bunch population.

\begin{figure}[!ht]
\center
\includegraphics[width=120mm]{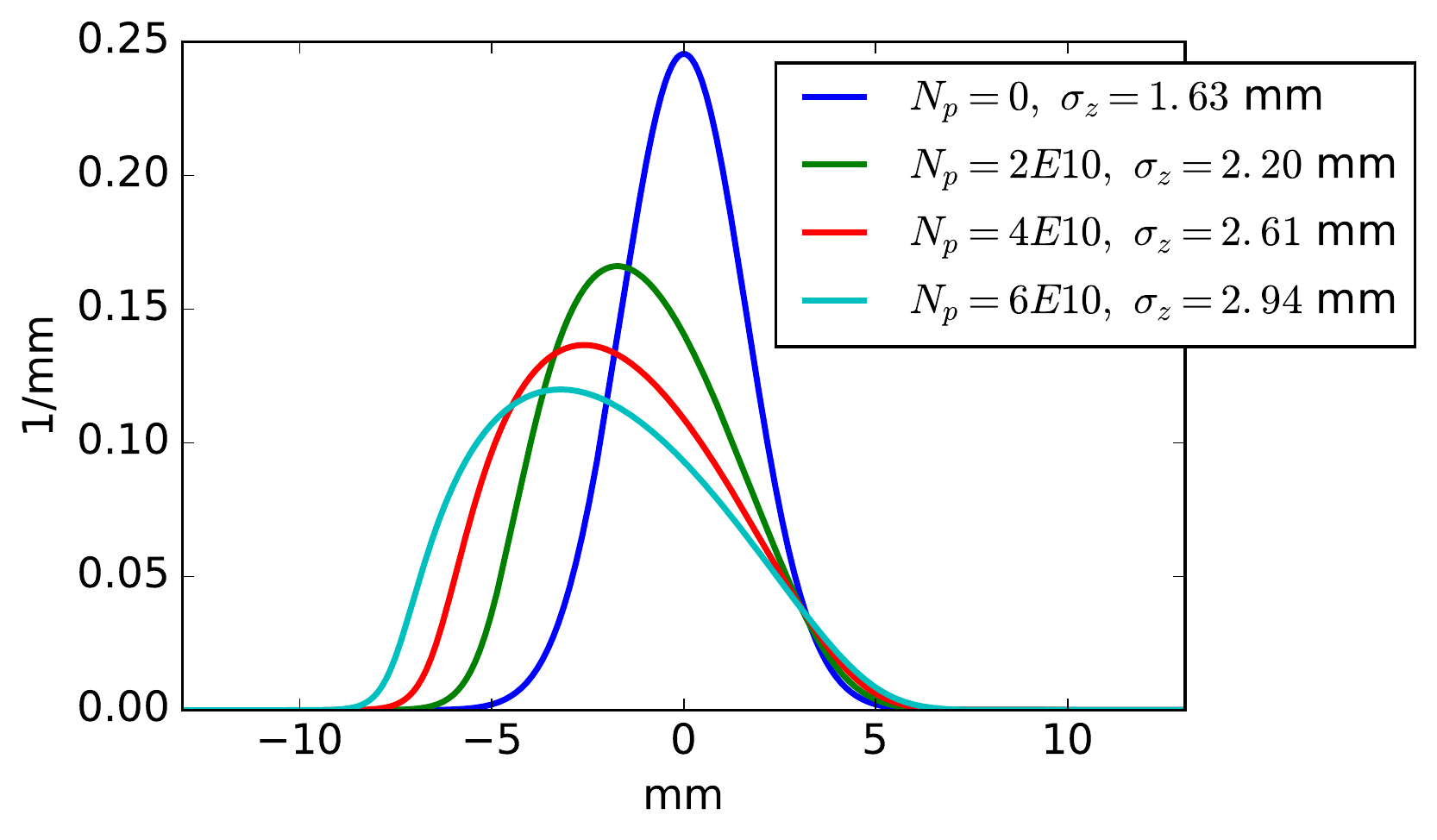}
   \caption{Longitudinal distribution for different bunch population as given by Ha\"\i ssinski equation.}
\label{haissinski}
\end{figure}

The bunch shapes for different bunch populations at the lowest energy of 45.6 GeV are shown in Fig.~\ref{haissinski}.

The bunch length is about 2.4 - 2.5 mm at the nominal current, but we have to remind that only the resistive wall effect has been taken into account for the moment. For the three shown bunch populations the tracking code gives exactly the same distribution. 

In Table~\ref{tab2} we have also reported the bunch length calculated by using the Ha\"\i ssinski equation with the nominal bunch population of Table~\ref{tab1} for all the four energies and due only to the resistive wall effect. We can see that the lowest energy case gives the largest effect, whilst for 120 and 175 GeV the lengthening due to the resistive wall wakefield is almost negligible.

\begin{table}[!ht]
\caption{Bunch length given by Ha\"\i ssinski equation with resistive wall impedance and with the parameters of Table~\ref{tab1}.}
\begin{center}
\begin{tabular}{|l|c|c|c|c|}
\hline
Beam energy (GeV)&  45.6 & 80 &  120 & 175 \\
\hline
Bunch length (mm)&  2.48 & 2.37 &  2.15 & 2.22 \\
\hline
\end{tabular}
\end{center}
\label{tab2}
\end{table}%

The potential well distortion theory described by the Ha\"\i ssinski equation predicts a bunch length increasing with current and a constant energy spread up to a given threshold, called microwave instability threshold, above which also the energy spread increases. In the microwave instability regime, even if the bunch is not lost, it could be characterized by internal turbulent motion which would compromise the machine performances. Several papers have been written to determine the microwave instability threshold~\cite{yunhai}.  In particular, in ref.~\cite{stupakov0}, the microwave instability due to the resistive wall wake fields was analyzed giving a criterion for the threshold evaluation. Applied to the FCC-ee case, it gives a threshold value of $N_p=8.1 \times 10^{10}$, a factor slightly higher than 2 with respect to the nominal bunch population.

\begin{figure}[!ht]
\center
\includegraphics[width=120mm]{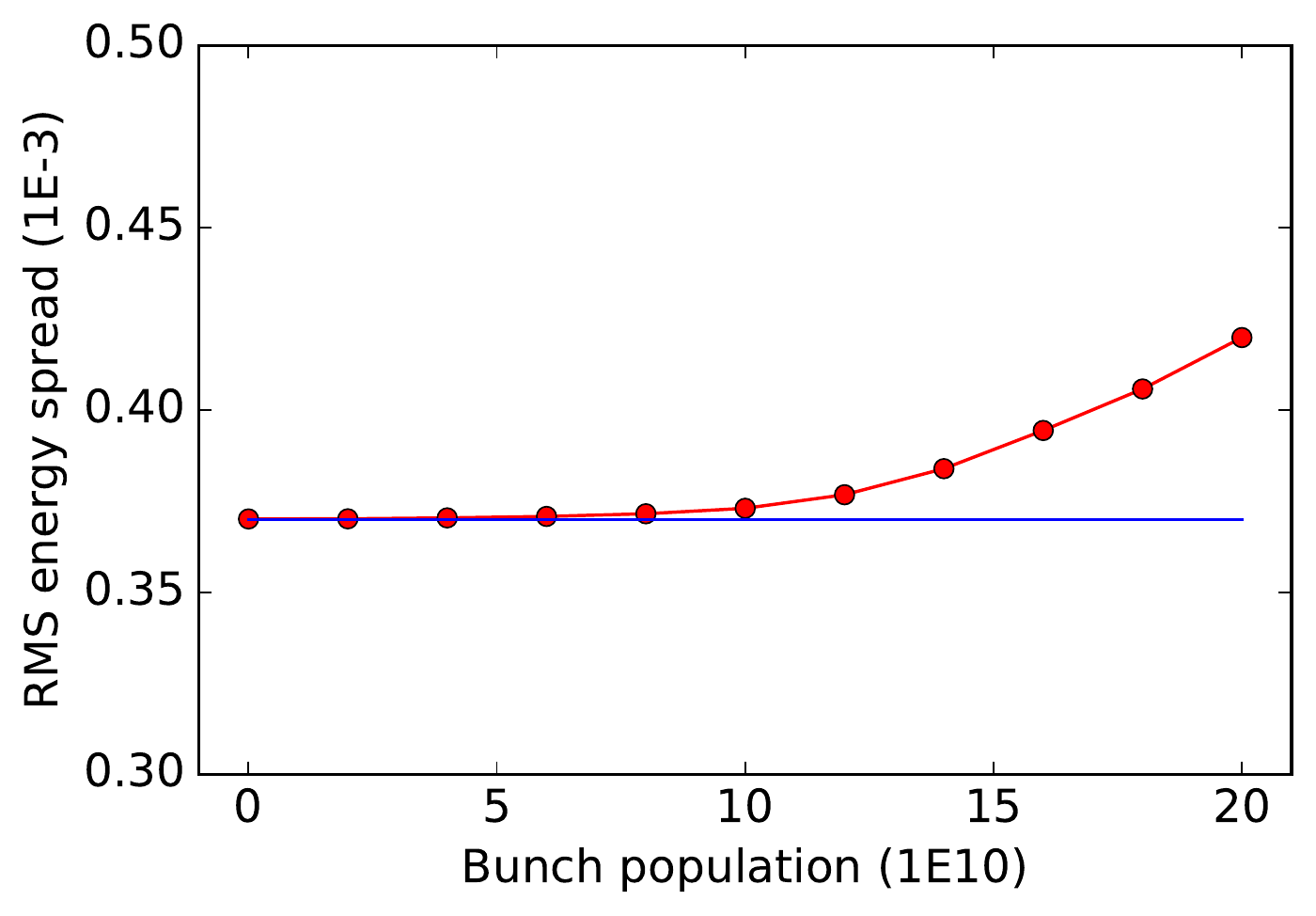}
   \caption{RMS energy spread as a function of bunch population given by the simulation code with only the resistive wall impedance.}
\label{energyspread}
\end{figure}

This value can be compared with the results of the tracking code. From Fig.~\ref{energyspread}, where we represented the RMS energy spread given by the code as a function of the bunch population, we can see that the energy spread starts to increase at about $8-10\times 10^{10}$. This is in a good agreement with the above analytical estimate.

As a further check of the tracking code results, a Vlasov-Fokker-Planck solver~\cite{warnok} has also been used, showing that up to a bunch population of $8\times 10^{10}$ the beam is stable and giving the onset of the instability at about $10-12\times 10^{10}$.

\begin{figure}[!ht]
\center
\includegraphics[width=120mm]{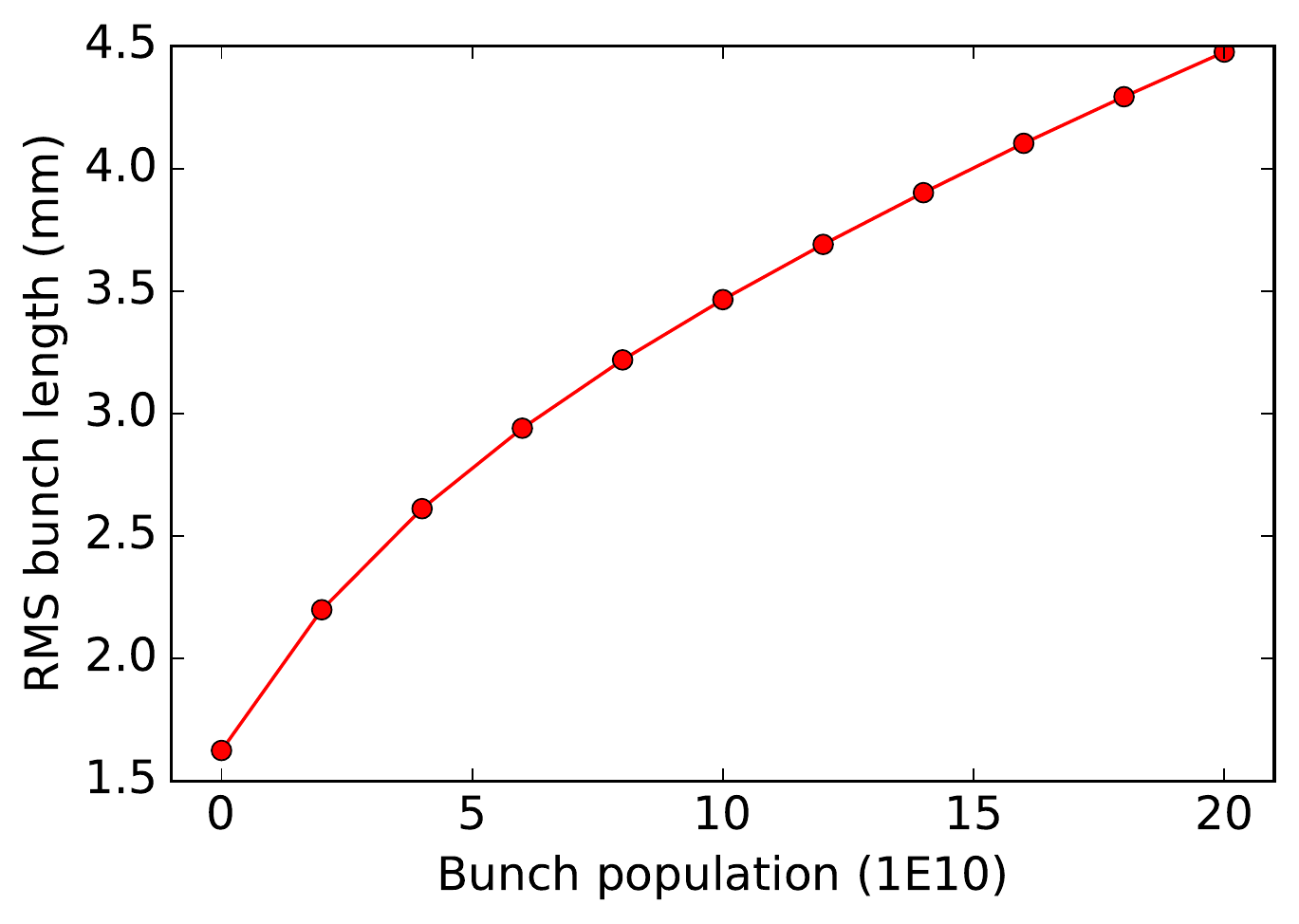}
   \caption{RMS bunch length as a function of bunch population as given by the simulation code with only the resistive wall impedance.}
\label{bunchlength}
\end{figure}

Finally, Fig.~\ref{bunchlength} shows the RMS bunch length, obtained with the simulation code, as a function of the bunch population up to an intensity of $2\times 10^{11}$, a value six times higher than the nominal bunch population.

\subsection{Multi-bunch effects}
\label{s:sub2}

A more critical situation is related to the transverse coupled bunch instability due to the long range transverse wake fields. In this case the study can be performed by considering the motion of the entire beam (not of the single bunch) as a sum of coherent oscillation modes, with coupled bunch modes to be taken into account. By considering the lowest azimuthal mode $m=0$ and a Gaussian bunch, the real part of the coupling impedance can produce stability or instability depending on the sign of the growth rate
\begin{equation}
\label{tcbi}
\alpha_{\mu,\perp}=-\frac{c I} {4 \pi (E_0/e) Q_{\beta}} \sum_{q=-\infty}^{\infty}Re \left[ Z_{\perp} \left( \omega_q \right) \right]
G_{\perp} \left(\sigma_{\tau} \omega'_q \right)
\end{equation}
where $I$ the total beam current, $Q_{\beta}$ the betatron tune, $\sigma_{\tau}$ the RMS bunch length in time, $G_{\perp}$ a form factor which, for our case, is about 1, and
\begin{equation}
\label{eq:modes}
\omega_q = \omega_0 \left(q N_b + \mu + Q_{\beta} \right) \qquad \omega'_q = \omega_q+\omega_0\xi \frac{Q_{\beta}}{\eta}
\end{equation}
with $N_b$ the number of bunches, $\xi$ the chromaticity, and $\omega_0$ the revolution frequency.

In the above equations, $\mu$ represents the $\mu^{th}$ coupled bunch mode, which goes from 0 to $N_b-1$. When $\alpha_{\mu}$ is positive, the corresponding mode is unstable. If we consider, as transverse impedance, the resistive wall one given by eq.~(\ref{rw2}), and ignore the term $-ib^4 \omega^2$, we observe that $Re \left[ Z_{\perp} \left( \omega \right) \right]$ depends on the sign of the frequency $\omega$. Negative frequencies produce unstable modes with an exponential growth given by eq.~(\ref{tcbi}), while positive ones give rise to damped oscillations. In addition to that, the resistive wall impedance grows approximately with the inverse of the square root of the frequency, determining the most dangerous coupled bunch mode when $\omega_q$ is as close to zero as possible. If we consider, as an example, the parameters given by table~\ref{tab1}, and $q=-1$, by denoting with $Q_0$ the integer part of the betatron tune, that is $Q_{\beta} = Q_0+ \nu_{\beta}$, with $\nu_{\beta}$ the fractional part of the tune, which plays a crucial role for this kind of instability, it comes out that the most dangerous coupled bunch mode is $\mu = N_b - Q_{0}-1=89949$, and this mode has its lowest negative frequency at $\omega_q=-\omega_0\left( 1-\nu_{\beta} \right)$. 

\begin{figure}[!ht]
\center
\includegraphics[width=165mm]{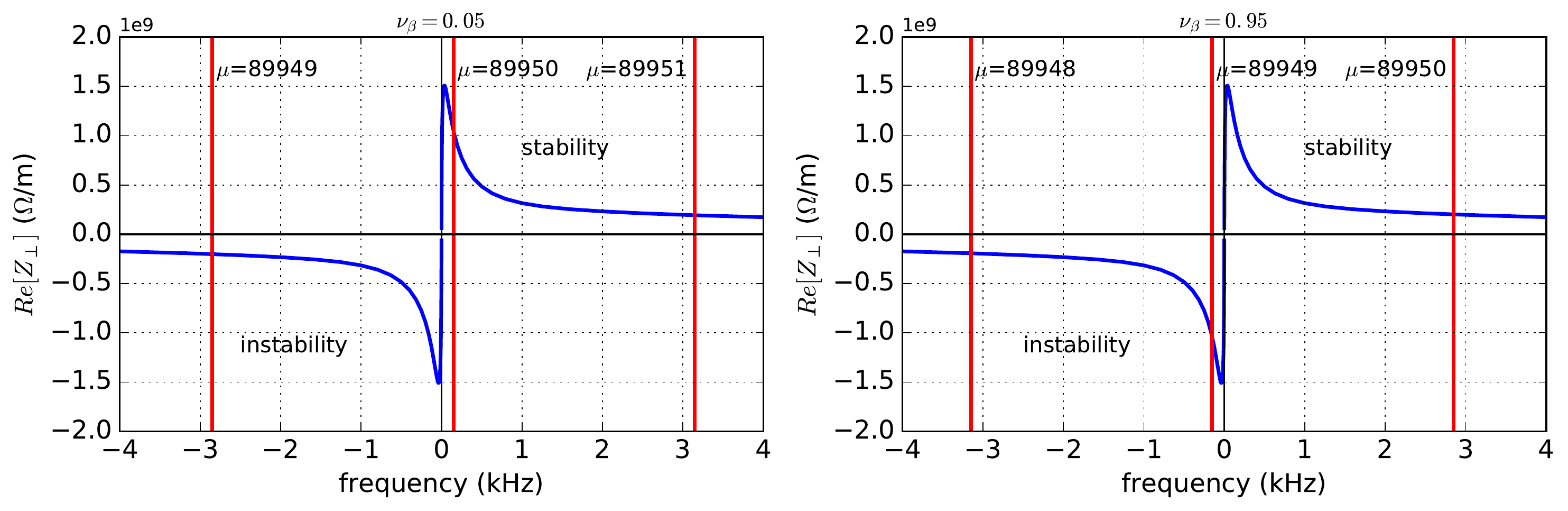}
  \caption{Coupled bunch spectrum and real part of the resistive wall impedance as a function of frequency around $f=0$ for fractional tune $\nu_{\beta}=0.05$ (left) and $\nu_{\beta}=0.95$ (right).}
\label{tbci1}
\end{figure}

Fig.~\ref{tbci1} shows the beam spectrum of three coupled bunch modes and the real part of the resistive wall impedance of a circular pipe of aluminium, with radius of 35 mm and three layers, close to zero frequency for two extreme cases of fractional part of the betatron tune, $\nu_{\beta}=0.05$ (left) and $\nu_{\beta}=0.95$ (right), and we see that a smaller fractional tune is preferred to alleviate the transverse coupled bunch instability because the impedance has a lower value. Due to dynamic aperture and  beam-beam issues, and since FCC-ee has 2 interaction points, the fractional tunes are indeed just above the integer~\cite{oide}, and therefore its fractional part is close to zero, mitigating the instability growth rate.

\begin{figure}[!ht]
\center
\includegraphics[width=165mm]{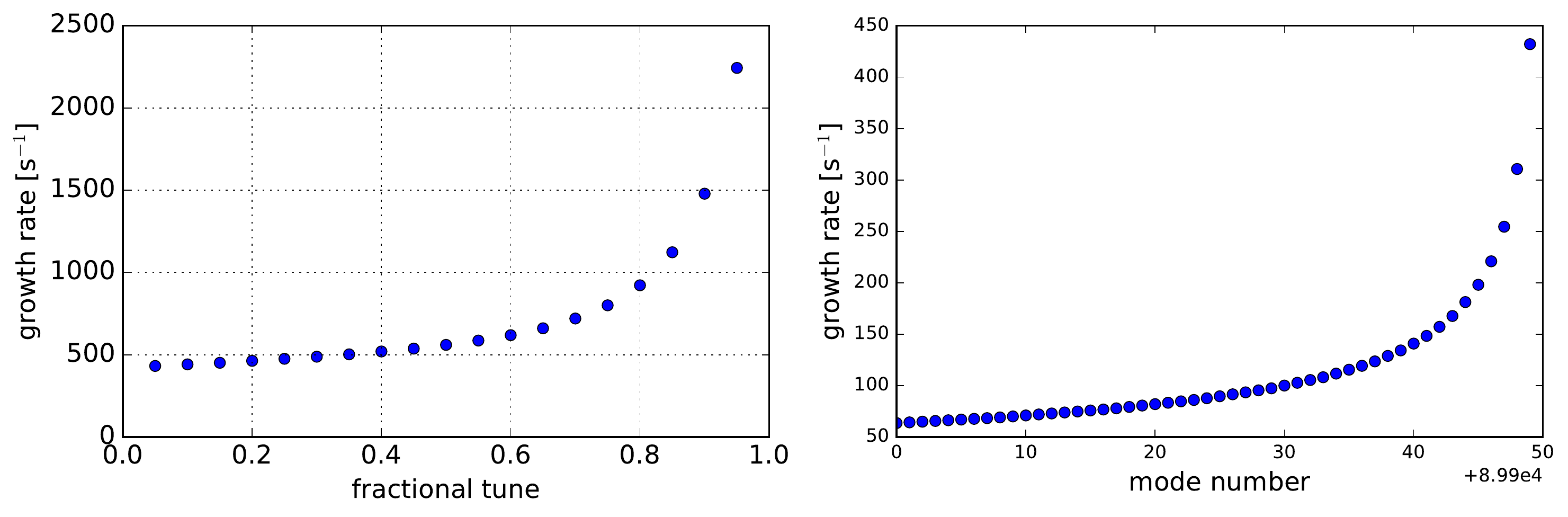}
  \caption{Growth rate as a function of the fractional tune for the most dangerous coupled bunch mode (left), and as a function of the coupled bunch mode number for the case with $\nu_{\beta}=0.05$ (right).}
\label{tbci2}
\end{figure}

If we consider, as an approximation, not a sum of the impedance over frequency in eq.~(\ref{tcbi}), but the coupling with a single betatron frequency line of the coupled bunch modes, the most dangerous unstable mode has a growth rate given approximately by
\begin{equation}
\alpha_{\perp}=\frac{c I} {4 \pi (E/e) Q_{\beta}} \frac{C}{2\pi b^3}\sqrt{\frac{CZ_0}{\pi |1-\nu_{\beta}|\sigma_c}}
\end{equation}
which, for the best case with $\nu_{\beta}=0.05$, gives a growth rate of about 432.4 s$^{-1}$, corresponding to a rise time of approximately 2.3 ms, that is about 7 machine turns. If the fractional tune increases, the situation worsens. In Fig.~\ref{tbci2} we show the growth rate as a function of the fractional tune for the worst coupled bunch mode (left plot), and as a function of the coupled bunch mode number for the case with $\nu_{\beta}=0.05$ (right plot).

A more precise calculation by considering the sum in eq.~(\ref{tcbi}) and by using the Laguerre polynomials with the DELPHI code confirms the values of the growth rates. As we can see from the right plot of Fig.~\ref{tbci2}, there are several unstable modes, and all of them need to be damped. In addition to that, even if the rise times are in the range of few milliseconds, which are not of particular concern for an accelerator machine, due to the large circumference, the rise times correspond to very few turns, making very challenging the realization of a feedback system. Some schemes that could deal with this problem have been proposed in ref.~\cite{drago}.

In addition to the previous instabilities, for the elliptic vacuum chambers, as we have seen in the previous section, there are  quadrupolar resistive wall wake fields that can produce a substantial tune variation with beam current in the multi-bunch regime. These are particularly important for the lowest energy FCC-ee option (45.6 GeV) with the highest beam current (1450 mA) distributed over a large number of circulating bunches.

For a rectangular vacuum chamber inside dipole magnets, the tune slope is estimated by~\cite{chao2}
\begin{equation}
\frac{d Q_{\beta}}{dI} =\pm \left( \frac{\pi r}{48 Q_{\beta}} \right) \left(\frac{Z_0}{E_0/e} \right) 
\left( \frac{R}{b}\right)^2 \left(\frac{L}{C} \right), \qquad r=1+\frac{b^2}{d^2}
\end{equation}
where $R$ is the radius of the machine, $b$ and $d$ the half height and half width of the vacuum chamber, $L/C$ the ring fraction with the rectangular chamber (dipoles).

This equation has reproduced very well the tune shifts observed in the high current lepton factories PEP-II~\cite{chao2} and DA$\Phi$NE~\cite{mikhail}. For the sake of estimates we consider the rectangular vacuum chamber with a cross section of 2bx2d= 70 mm x 120 mm and assume that the dipole vacuum chambers occupy $L/C=0.59$ of the ring, similarly to the HER of PEP-II. In these conditions, for the energy of 45.6 GeV we obtain the tune variation of 
\begin{equation}
\frac{d Q_{\beta}}{dI} =\pm  \frac{0.253}{A} 
\end{equation}
As a consequence, during the beam injection and up to the nominal value of 1.45 A, the tune variation would be as large as about 0.37. In order to cope with such big tune changes additional feedback systems would be necessary to keep the tune constant as required by the beam-beam interaction, dynamic aperture considerations and other beam dynamics issues. 

For what concerns possible longitudinal coupled bunch instabilities excited by HOMs, at this stage it is not possible to quantify their impedance contribution, but we can estimate the maximum shunt impedance that gives a growth rate that can be compensated by the natural radiation damping.

\begin{figure}[!ht]
\center
\includegraphics[width=115mm]{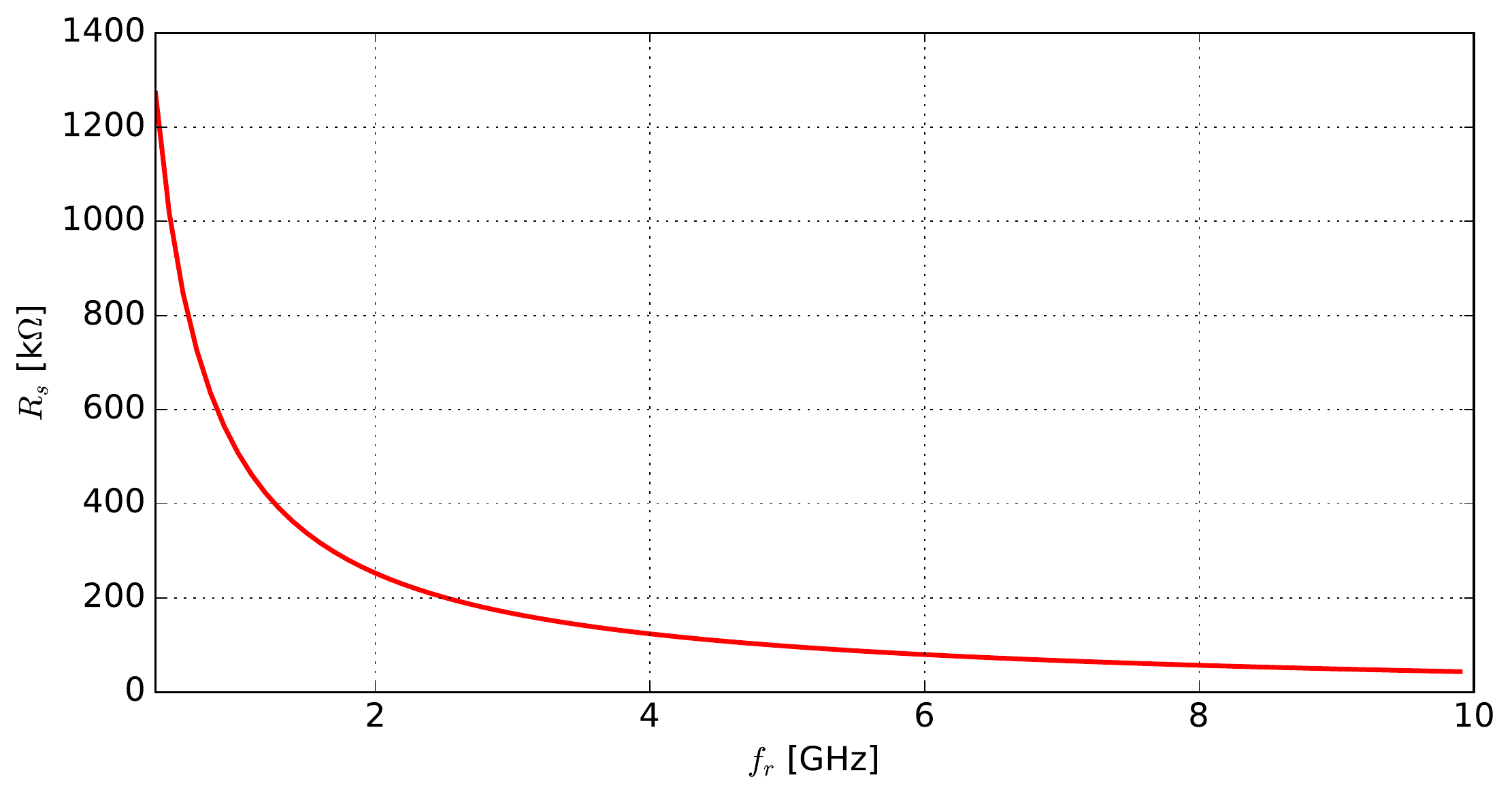}
 \caption{Maximum shunt impedance of a HOM as a function of its resonance frequency, producing a growth rate that is compensated by the natural radiation damping.}
\label{HOMs}
\end{figure}

Similarly to the transverse case, starting from the linearized Vlasov equation, and developing any perturbation of the beam distribution as sum of coherent oscillation modes, it is possible to obtain an eigenvalue system representing the coherent frequencies of the modes. By neglecting the coupling between the different azimuthal modes, and by considering only the lowest longitudinal azimuthal mode $m = 1$, it is possible to show that the real part of the HOM impedance can produce stability or instability depending on the sign of the growth rate
\begin{equation}
\alpha_{\mu,||}=\frac{\eta I}{4 \pi \left(E_0/e \right) Q_s} \sum_{q=-\infty}^{\infty} \omega_q Re \left[ Z_{\||} \left( \omega_q \right) \right] G_{||} \left(\sigma_{\tau} \omega_q \right)
\end{equation}
with $Q_s$ the synchrotron tune and $\omega_q = \omega_0(qN_b+\mu+Q_s)$. Stability in this case occurs for negative frequencies because the real part of the longitudinal impedance is always positive, and the worst and simplest unstable case is when the HOM has its resonant angular frequency $\omega_r$ equal to $\omega_q>0$. Similarly to the transverse case, if we consider, as an approximation, not a sum of the impedance over frequency, but the coupling with a single synchrotron frequency line of the coupled bunch modes, the most dangerous unstable mode has a growth rate given approximately by
\begin{equation}
\label{eq:growth_long}
\alpha_{||}=\frac{\eta I}{4 \pi \left(E_0/e \right) Q_s} \omega_r R_s
\end{equation}
with $R_s$ the HOM shunt impedance. Also in this case $G_{||}(x) \simeq 1$, if $f_r \ll 25$ GHz. This growth rate has to be compared with the natural damping rate due to the synchrotron radiation, which, for the lowest energy machine, is about 1320 turns. In Fig.~\ref{HOMs}, we have represented the maximum HOM shunt impedance of eq.~(\ref{eq:growth_long}) as a function of the resonance frequency, such that the corresponding growth rate is exactly balanced by the radiation damping. Of course, similarly to the transverse case, also here a feedback system has to be developed as a further safety knob.

\section{Other important impedance sources}
\label{s:sec3}

In the previous section, by discussing the effects of the resistive wall, we have seen that its contribution on the beam dynamics is very important, requiring, in some cases, active feedback systems to keep under control beam instabilities.
Due to the above results, we can consider, as a rough indication to accept the impedance of a given device in the design stage, the comparison of its wake potential with the resistive wall one.

Generally, for a high energy e+e- machine, the handling of synchrotron radiation represents one of the most challenging tasks. As a consequence of this, it is necessary to have a sufficient number of cavities to recover the lost energy, and, in addition to that, synchrotron radiation absorbers, necessary to cope with extra heating and eventual background. These devices  could be important sources of impedance, and their evaluation is considered in this section and compared with the resistive wall contribution.

\begin{figure}[!ht]
\center
\includegraphics[width=115mm]{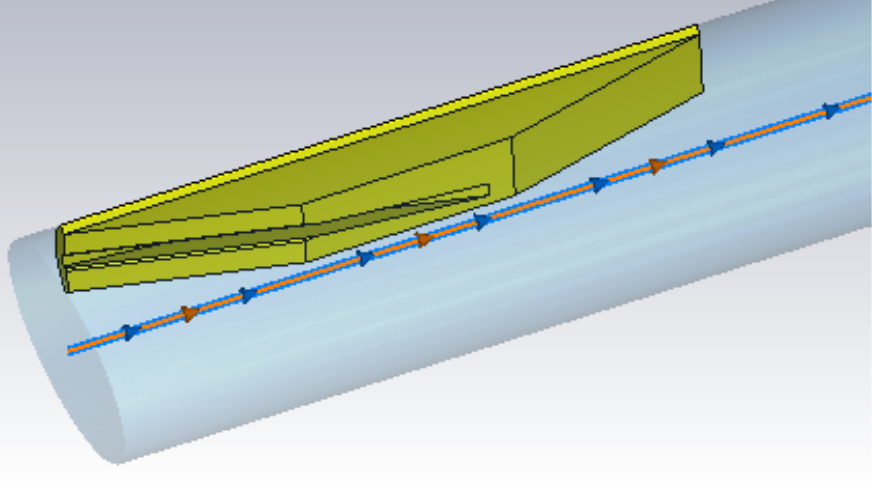}
  \caption{Elliptic vacuum chamber with one absorber~\cite{kersevan}.}
\label{old_absorber}
\end{figure}

Let us first estimate the impact of the synchrotron radiation absorbers. For FCC-ee a synchrotron radiation absorber will be installed every 4-6 meters, with the purpose of intercepting the radiation that, otherwise, would impact on the beam chamber. Due to their large number, the absorbers represent a very important source of machine impedance.  Initially, an elliptical vacuum pipe, with absorbers on one side of the major semi-axis, as shown in Fig.~\ref{old_absorber}, was proposed. A total of 9228 absorbers was estimated~\cite{kersevan}.

\begin{figure}[!ht]
\center
\includegraphics[width=115mm]{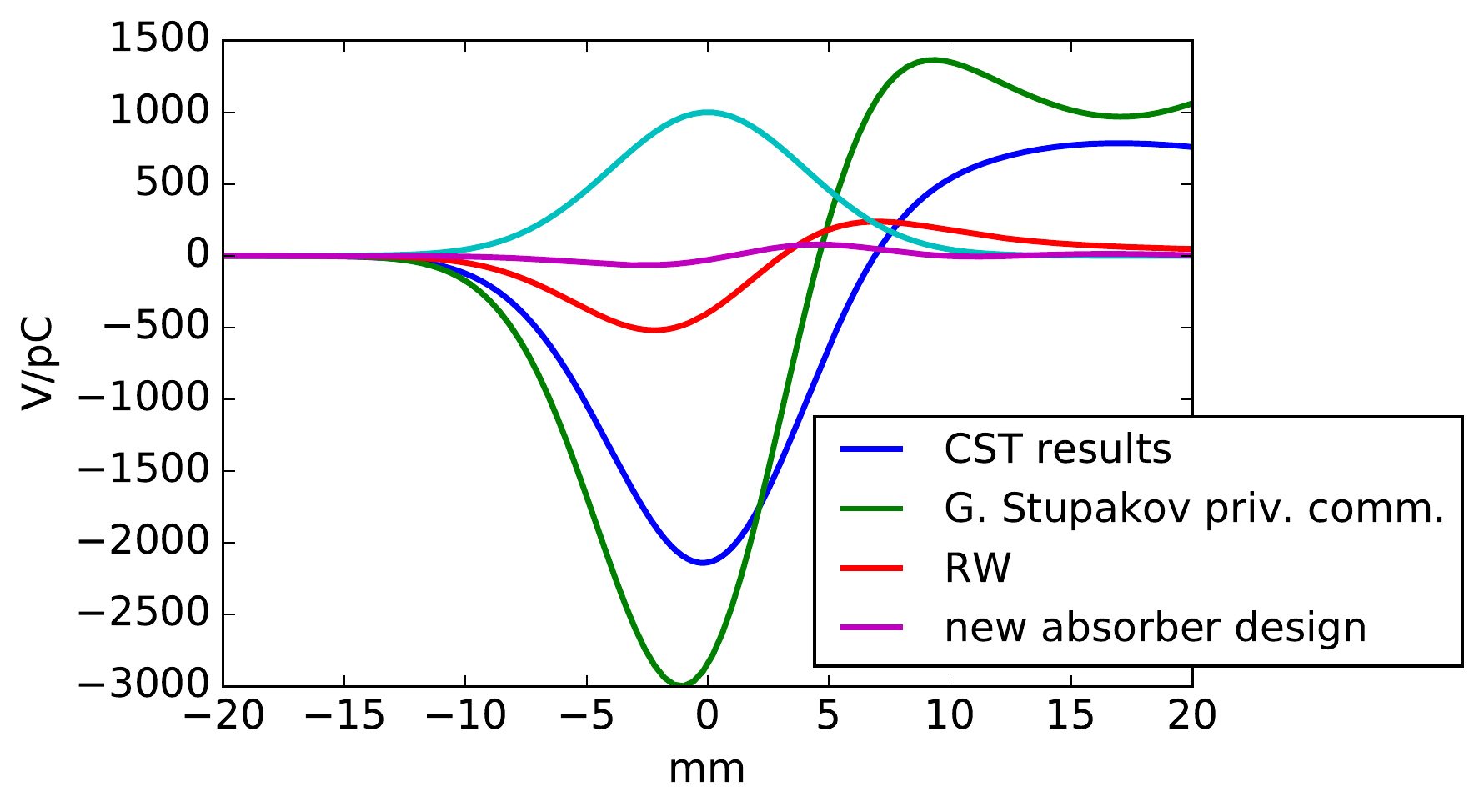}
  \caption{Longitudinal wake potential of a 4 mm Gaussian bunch produced by: resistive wall (red line), 9228 absorbers (CST simulations, blue line), 18456 absorbers (two absorbers at each side) with a more approximated approach taking into account a rectangular geometry~\cite{stupakov} (green line), 10000 absorbers for a new alternative design (CST simulation, magenta line).}
\label{wp_old_abs}
\end{figure}

Of course, the contribution of a single absorber could be neglected, but the effect of all of them is expected to be prohibitively large compared to the resistive wall of a circular beam pipe. In Fig.~\ref{wp_old_abs} we have represented the wake potential of a 4 mm Gaussian bunch given by all the absorbers as obtained by CST Particle studio~\cite{CST} and by using a simpler and more approximated approach that takes into account a rectangular geometry with two absorbers at each side~\cite{stupakov}. In the same figure the wake potential of 10000 absorbers given by CST considering a new alternative design described below is also shown.

Even if there is a certain degree of uncertainty in the results, and clearly the more approximated approach may overestimate the longitudinal wake potential using two absorbers for symmetry reasons, both approaches demonstrate that the impedance contribution of the absorbers in this condition is unacceptable. In addition to this result, if we further consider the previous comments on the tune shifts produced by the quadrupolar wake fields of the elliptic vacuum chamber, it is clear that the elliptic cross section is not the best solution for FCC-ee from the impedance point of view.

A proposed alternative is the installation of a circular vacuum chamber, with radius 35~mm and winglets on both sides, as the one of Super-KEKB~\cite{SuperKEKB}. Bellows with RF fingers and valves are to be designed with the same profile. Transitions to a circular cross section chamber (without winglets) should be necessary only to connect the pipe to the RF system.

The absorbers are metallic devices shaped like a trapezoid, with a total length of 30 cm, and they are inserted inside the chamber winglets, at about 42.5 mm from the beam axis. Placing slots for vacuum pumps just in front of the absorber allows to efficiently capture the synchrotron radiation and the molecule desorption. The pumping slots have a racetrack profile, length of 100-120~mm and width of 4-6~mm. Behind the slots, a cylindrical volume and a flange will be installed to support a NEG pump~\cite{kersevan}.

\begin{figure}[!ht]
\center
\includegraphics[width=115mm]{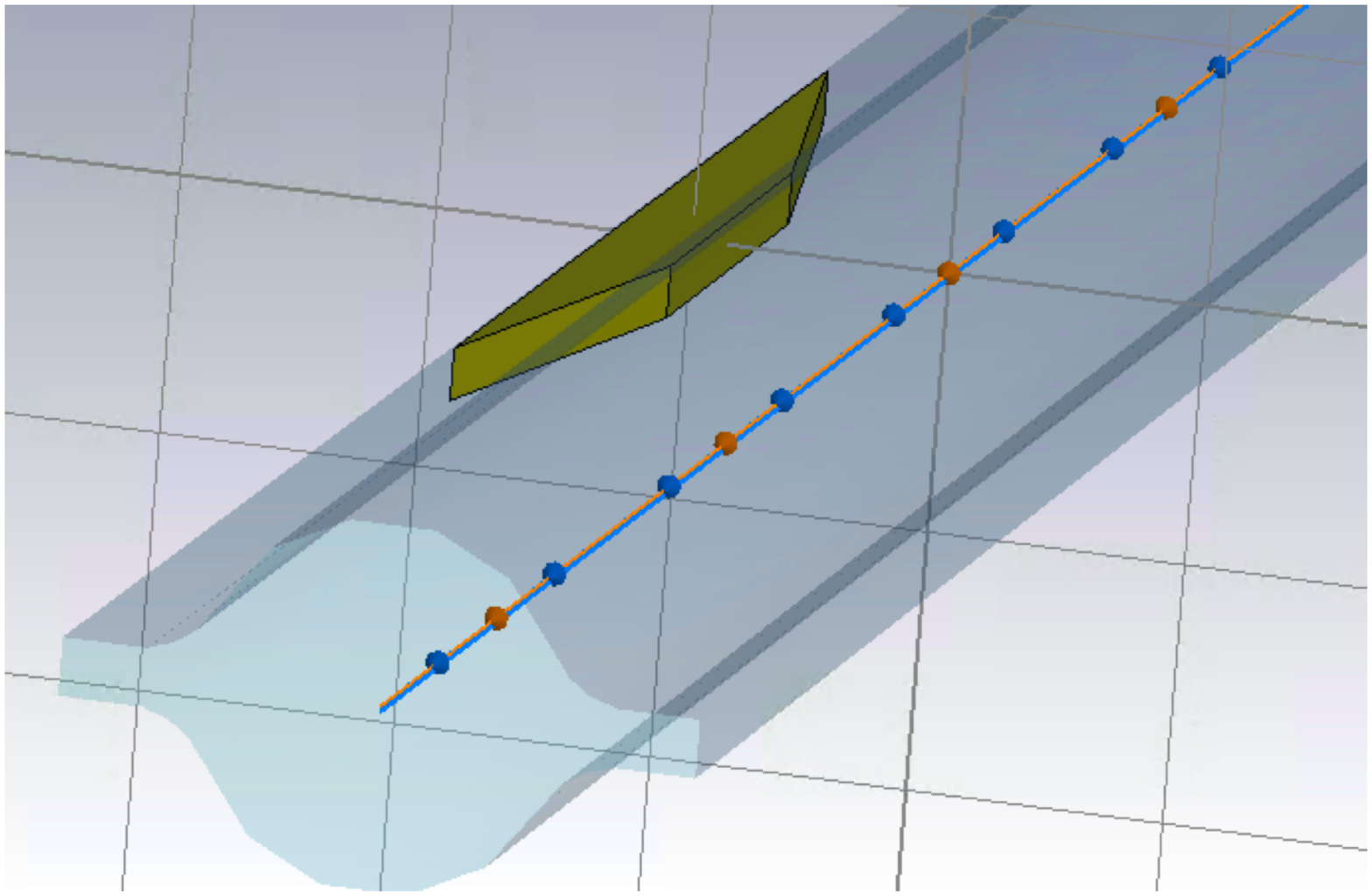}
  \caption{3D model of the FCC-ee vacuum chamber with winglets and a synchrotron radiation absorber used for CST simulations.}
\label{Absorber}
\end{figure}

Impedance studies of the beam chamber profile with one absorber insertion have been performed using CST. In Fig.~\ref{Absorber}, the geometry of the FCC-ee beam chamber used in CST simulations is shown together with a detail of the absorber inside the beam chamber. Pumping slots and pumps are not included in this preliminary model.
Simulations show that below about 3~GHz the longitudinal impedance is purely inductive, giving, for 10000 elements, a longitudinal broadband impedance $Z/n$ of about 1~m$\Omega$.

\begin{figure}[!ht]
\center
\includegraphics[width=115mm]{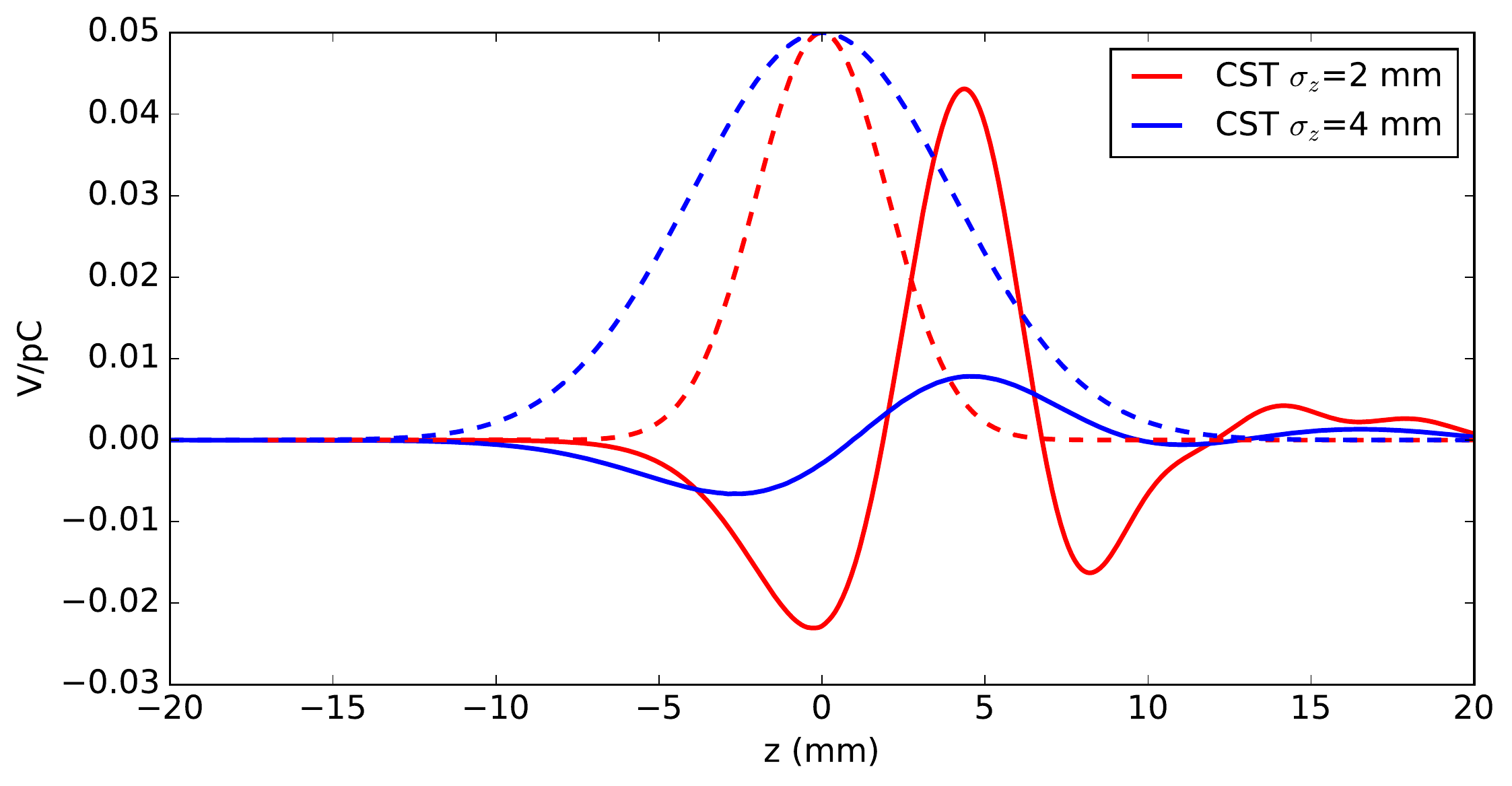}
  \caption{Wake potential of a single abosorber for 2 and 4 mm RMS bunch length from CST code.}
\label{wpabsorbers}
\end{figure}

In Fig.~\ref{wpabsorbers}, the wake potential of a 2 and 4 mm bunch length is shown. Even if further analysis is needed, and this first evaluation could overestimate the impedance, by multiplying this wake by the number of absorbers, about 10000 elements, we see that their contribution is much lower than the resistive wall one, as shown in Fig.~\ref{wp_old_abs}. About the transverse contribution to the impedance of a single absorber, this is so low that, up to now, we did not manage to obtain reliable results.

\begin{figure}[!ht]
\center
\includegraphics[width=80mm]{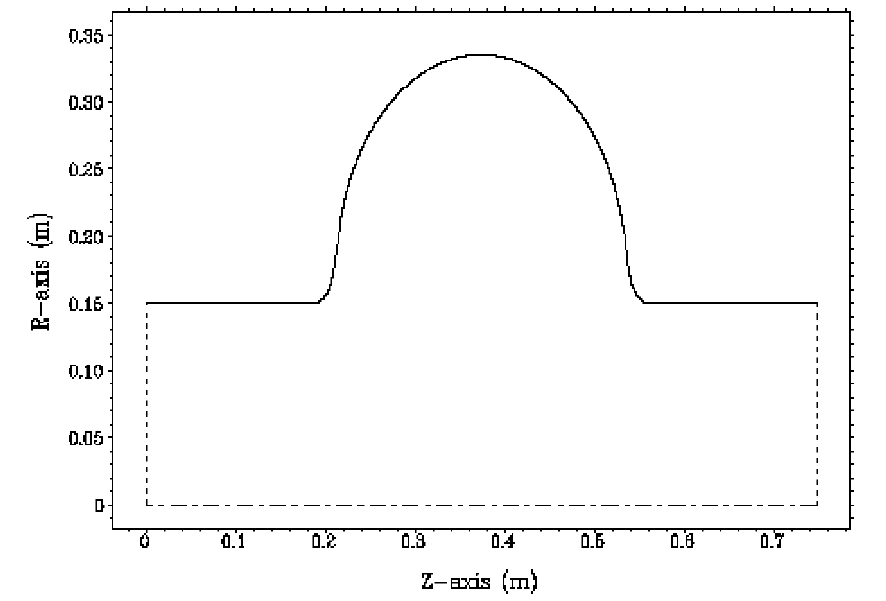}
\includegraphics[width=80mm]{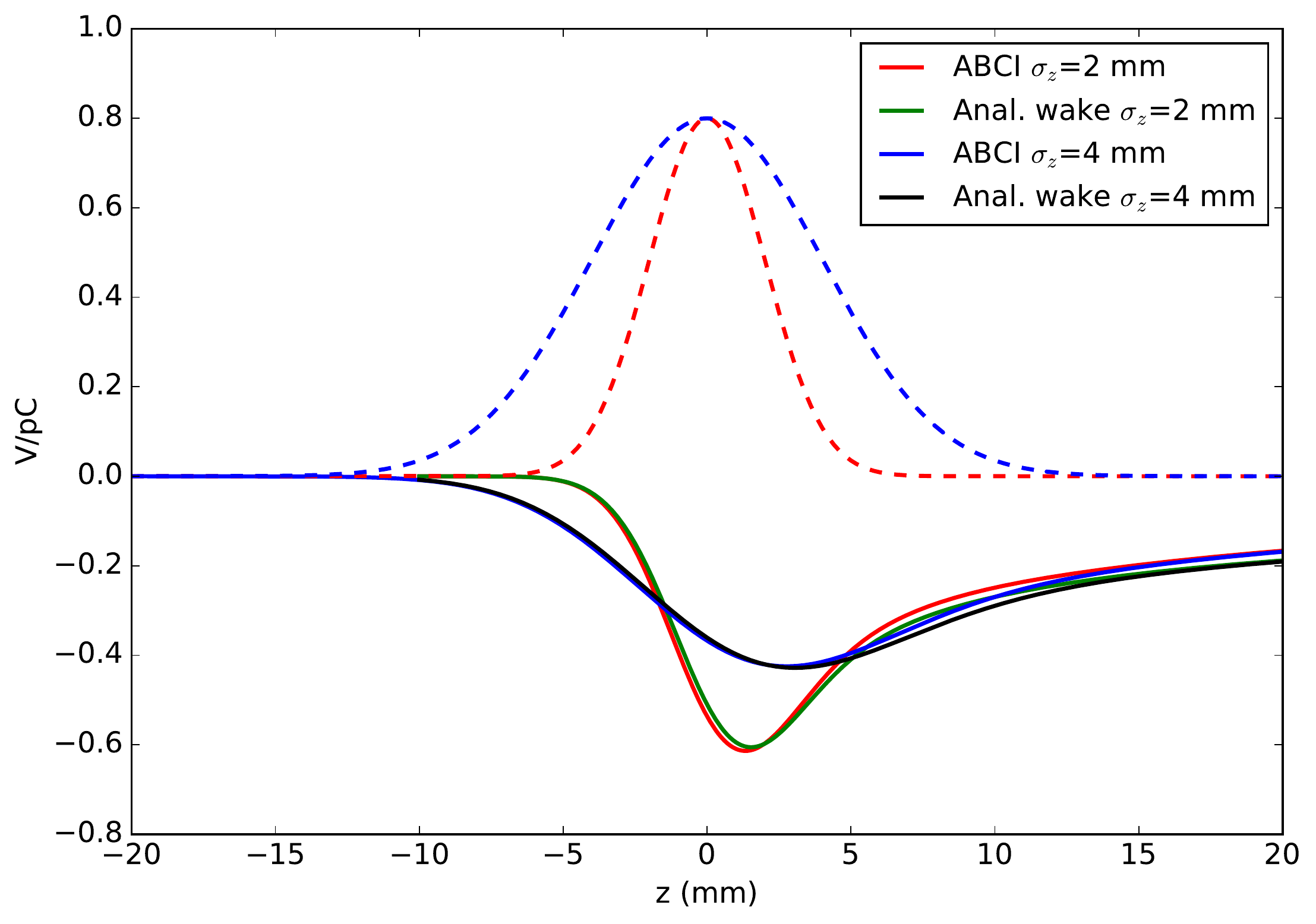}
  \caption{400 MHz single cell cavity used in ABCI (left) and wake potentials for 2 and 4 mm RMS bunch length, as given by the ABCI code (red and blue lines), and by eq.~\ref{eq:cavity} (green and black).}
\label{cavity}
\end{figure}

Another important source of impedance is represented by the RF system. Several options are under investigation~\cite{calaga}, and in our study 400 MHz cavities with a single cell, as the one shown in Fig.~\ref{cavity} left side, have been considered. The wake potentials for 2 and 4 mm RMS bunch length, as given by the ABCI code~\cite{ABCI}, are shown on the right side of the same figure. 

In order to check the ABCI results, in the same plot we have also represented the analytical wake potentials, for the same bunch lengths, of a cavity with attached tubes at high frequencies, given by the expression~\cite{palumbo}
\begin{equation}
\label{eq:cavity}
W(x)=\tilde{W} |x|^{1/4}e^{-x}\left[I_{-1/4}(x)+\textrm{sign}(z)I_{1/4}(x) \right]
\end{equation}
with $x=[z/(2\sigma_z)]^2$. As we can see from the figure, the analytical curves are almost  superimposed to those given by ABCI.

\begin{figure}[!ht]
\center
\includegraphics[width=120mm]{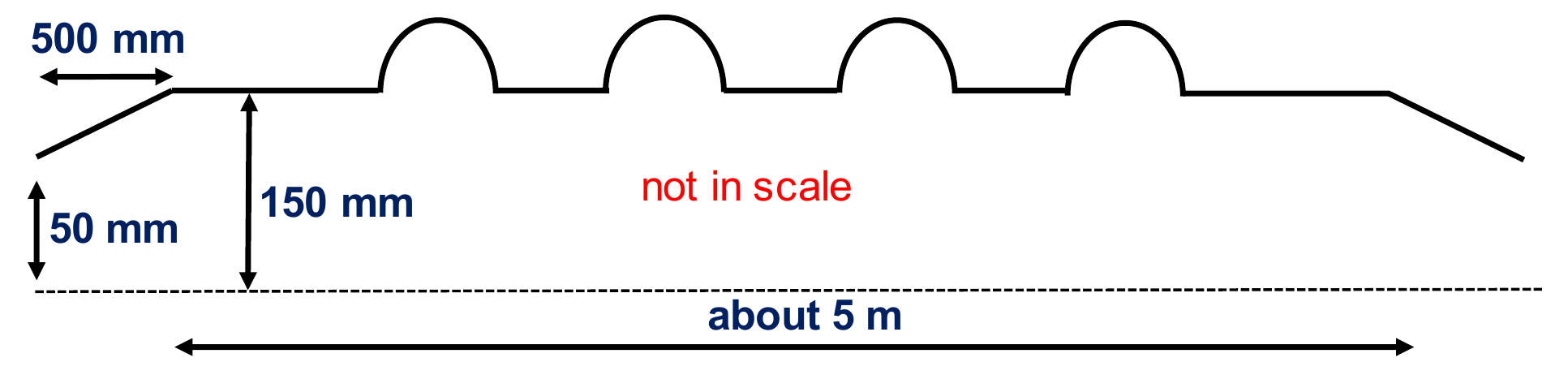}
\caption{Sketch, not in scale, of a group of 4 cells with tapers at each end.}
\label{rfsystem}
\end{figure}

By using the single cell, a group of 4 cavities are put together as shown in Fig.~\ref{rfsystem}, and connected to other two groups by two tapers. A total number of 100 cells has been considered as a global contribution to the impedance model, and, as a consequence, 25 double tapers. Of course the impedance produced by the tapers strongly depends on their length, which we have considered here to be of 500 mm.

\begin{figure}[!ht]
\center
\includegraphics[width=120mm]{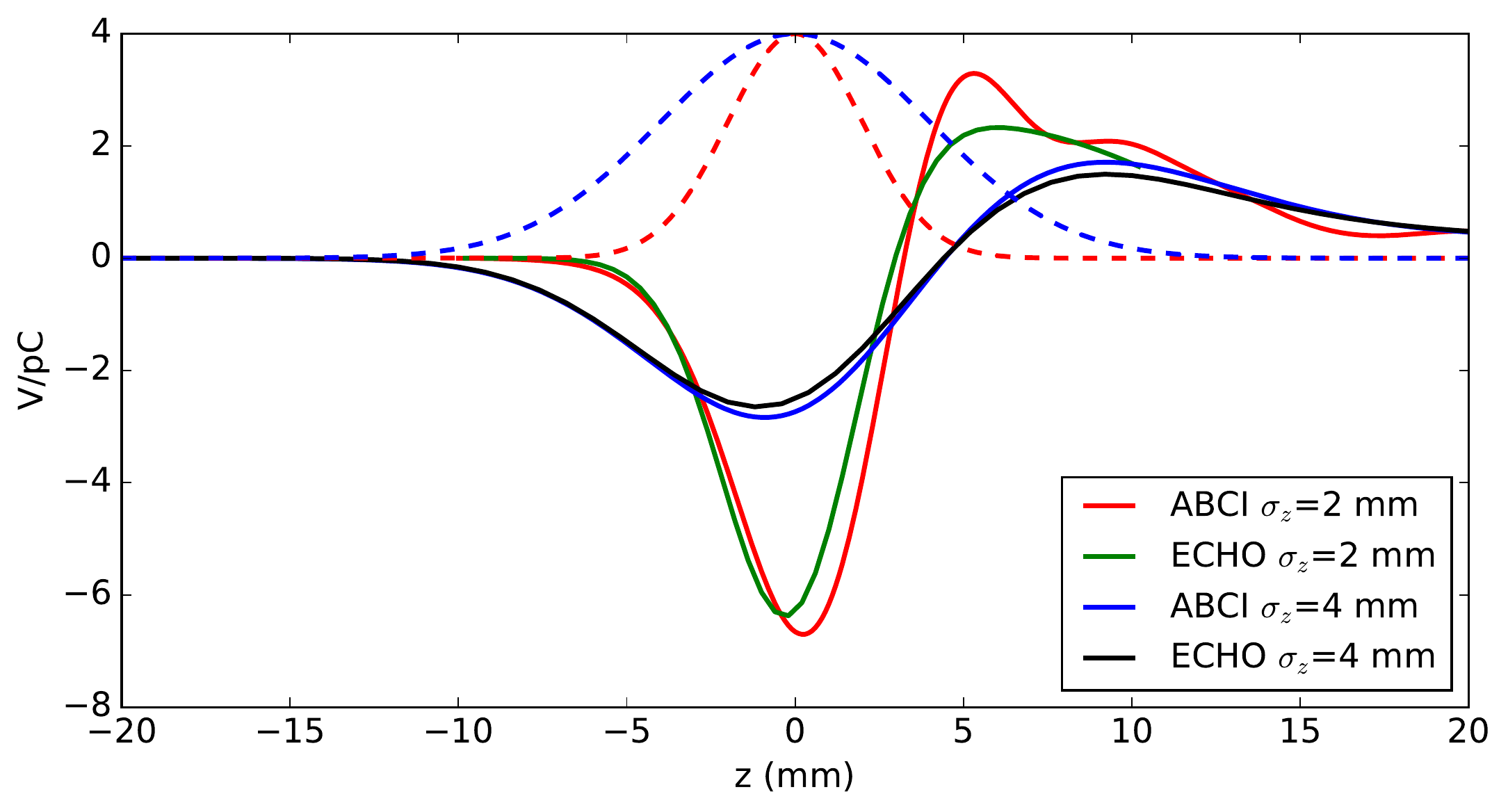}
\caption{Wake potentials of a double taper for 2 and 4 mm RMS bunch length, as given by the ABCI code (red and blue lines), and by the ECHO code (green and black).}
\label{tapers}
\end{figure}

The wake potential of a single double taper (in and out, considered independent) in this condition is shown in Fig.~\ref{tapers}. The results have been obtained with ABCI and another electromagnetic code, ECHO~\cite{echo,stupakov1}, and the agreement can be considered satisfactory.

\begin{figure}[!ht]
\center
\includegraphics[width=120mm]{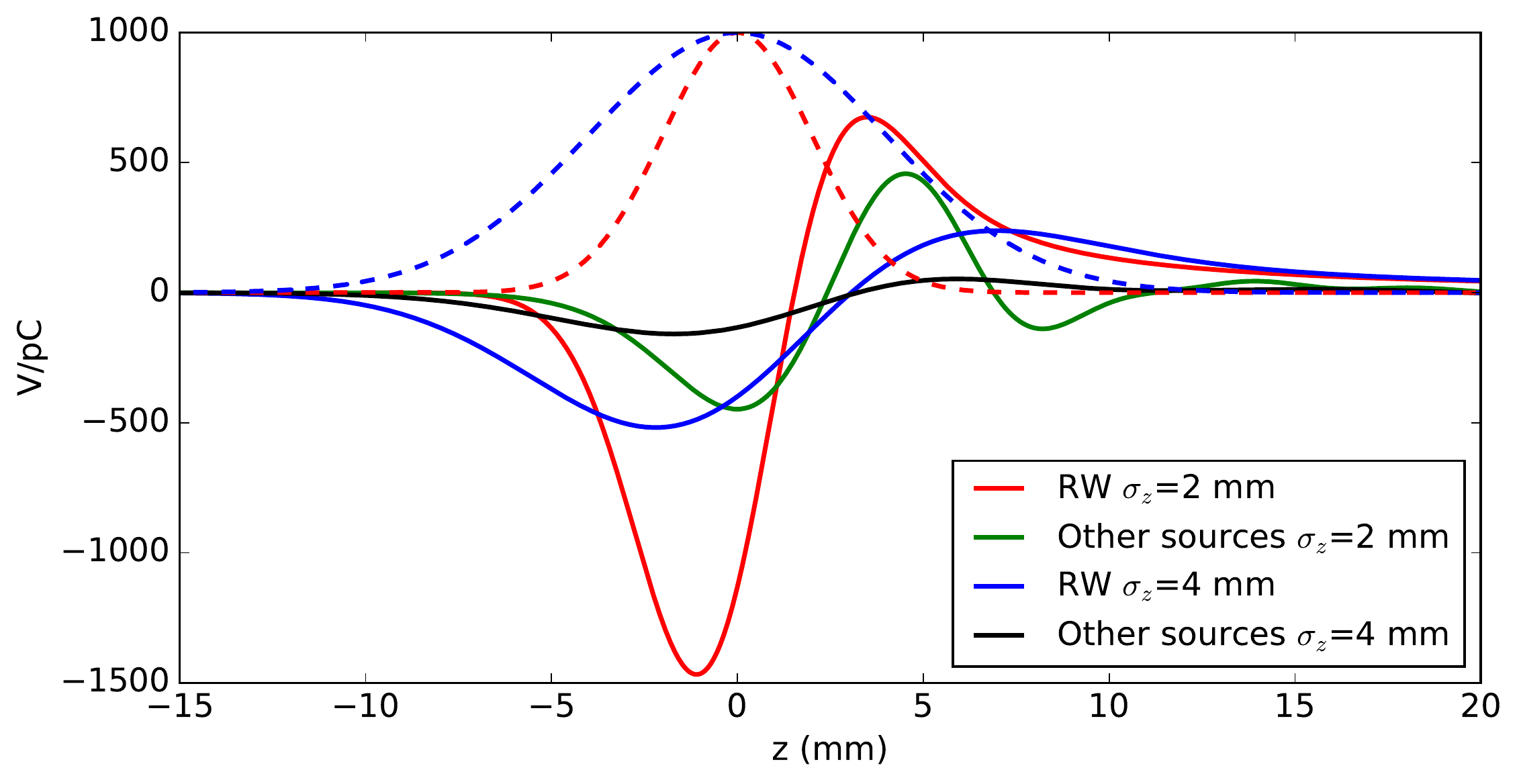}
\caption{Total wake potentials for 2 and 4 mm RMS bunch length due to the RF system and the absorbers (green and black lines) compared with the resistive wall contribution (red and blue lines).}
\label{waketot}
\end{figure}

Since the RF system (including the connecting tapers) and the absorbers are very important sources of impedance, in particular these last ones for their high number, in Fig.~\ref{waketot} we have represented the total longitudinal wake potential of these sources for 2 and 4 mm bunch length compared with the resistive wall contribution. In the figure the red and the blue curves have to be compared with the green and the black one respectively. We can see a factor of about 3 between the resistive wall contribution and the other sources, indicating that, up to now, the resistive wall remains the main source of impedance and wake fields.

Of course, also the transverse contribution of the previous devices has to be taken into account. However, since their longitudinal contribution is quite smaller than the resistive wall, even if a more careful evaluation has to be carried out, we do not expect surprising results. There are several other sources of impedance, such as bellows, RF fingers, BPMs and other devices for diagnostics, and the criterion for their design should be such as to generate an impedance much smaller with respect to the resistive wall part.

Possible trapped modes in the interaction region deserve special studies, and work on other collective effects, such as the fast ion and the electron cloud instabilities, is in progress.

\section{Conclusions and outlook}

In this paper we have discussed single beam collective effects in FCC-ee due to the beam coupling impedance. In particular we focused our study primarily on the resistive wall effects because this is, up to now, the main source of impedance. 

We have found that, in the single bunch case, the transverse mode coupling instability is about a factor 6 higher than the nominal bunch population at the lowest energy (45.6 GeV), and even higher for the other machine energies. Also the microwave instability due to the resistive wall has a margin of safety of about 2.4 with respect to the bunch population. However these numbers will be reduced as other sources of impedance will be taken into account.

Regarding the multi-bunch effect, we have concluded that the resistive wall transverse coupled bunch instability has to be counteracted by a feedback system, which requires innovative ideas for its design. Also the quadrupolar resistive wall wake fields, due to the elliptic vacuum chamber, produce a very high tune variation, which would not exist for the circular geometry. For the longitudinal case, at this stage, it is not possible to evaluate the characteristics of trapped HOMs, but an estimate of the maximum allowed shunt impedance as a function of the resonant frequency has been given.

In addition to the assessment of the resistive wall effects, we have started the evaluation of the impedance budget for other devices, with the goal of designing them in order to give an impact on the beam dynamics much smaller than that of the resistive wall itself. With an accelerator of 100 km of length, this is a long work, and the strategy is to identify the most important sources of impedance. We have started with the synchrotron radiation absorbers and the RF system. For the former, it resulted that the initial proposal could not be accepted, and, consequently, a new geometry of the beam pipe has been presented. Both contributions together amount to about one third of the resistive wall one in the longitudinal plane. Their evaluation in the transverse plane is in progress.

Other machine devices are going to be studied with the goal of comparing their contribution to the resistive wall impedance. Also two-stream instabilities, such as the fast ion and the electron cloud, must be determined.

\section*{Acknowledgements}

We acknowledge many helpful and stimulating discussions with R. Calaga, R. Kersevan, K. Oide, E. Shaposhnikova, G. Stupakov, D. Zhou, and F. Zimmermann.

%\begin{thebibliography}{9}   % Use for  1-9  references

\end{document}